\documentclass{ws-book9x6}
\usepackage{ws-book-thm}  
\usepackage{ws-book-har}  
\usepackage{bm,mathrsfs,isotope,color,aas_macros}
\usepackage[pdfpagelabels=false,colorlinks=true,allcolors=black]{hyperref}

\newcommand{\vectau}{{\bm \tau}}
\newcommand{\vecrho}{{\bm \rho}}

\begin{document}

\chapter[Hyperonization in compact stars]{Hyperonization in compact stars}\label{ch1}

\author{Armen Sedrakian}
\address{Frankfurt Institute for Advanced Studies, \\Ruth-Moufang-Str. 1, 60438 Frankfurt am Main, Germany\\
  Institute of Theoretical Physics,  University of Wroc\l{}aw, \\pl. M. Borna 9, 50-204 Wroc\l{}aw, Poland}

\author{Jia-Jie Li}
\address{School of Physical Science and Technology, Southwest University, Chongqing 400700, China}

\author{Fridolin Weber}
\address{Department of Physics, San Diego State University, \\San Diego, CA 92182, USA\\
  Center for Astrophysics and Space Sciences,
   University of California at San Diego, La Jolla, CA 92093, USA
}

%\begin{abstract}
  We review the covariant density functional approach to the equation of state of the dense nuclear matter in compact stars. The main emphasis is on the hyperonization of the dense matter, and the role played by the
  $\Delta$-resonances. The implications of hyperonization for the astrophysics of compact stars, including the equation of state, composition, and stellar parameters are examined. The mass-radius relation and tidal deformabilities of static and rapidly rotating (Keplerian) configurations are discussed in some detail. We briefly touch upon some other recent developments involving hyperonization in hot hypernuclear matter at high- and low-densities.
%\end{abstract}

\section{Introduction}
\label{sec:Intro}

Compact (or neutron) stars represent the endpoints of the evolution of ordinary stars.  They provide a natural astrophysical laboratory for particle and nuclear physics under conditions that are largely different from and often inaccessible in terrestrial laboratories.  For example, densities reached in compact stars are by a factor 5-10 larger than found in ordinary nuclei. In the 60s and 70s of the past century it has been conjectured that they may contain non-nucleonic constituents of matter, for example, hyperons~\cite{Ambartsumyan1960SvA,Ambartsumyan1961AZh,Leung1971ApJ,Pandharipande1971NuPhA,Moszkowski1974PhRvD} or deconfined quark matter~\cite{Itoh1970PThPh,Collins1975PhRvL}. The state of (hyper)nuclear matter found in compact stars may feature some extraordinary facets, such as superfluidity and superconductivity, a trapped neutrino component at the early stages of evolution, and super-strong magnetic fields, for reviews see~\citet{Shapiro:1983du,Glendenning2012compact,weber1999pulsars,Sedrakian2007PrPNP,Oertel2017}.

\begin{figure}[t] \centerline{\includegraphics[width=11cm]{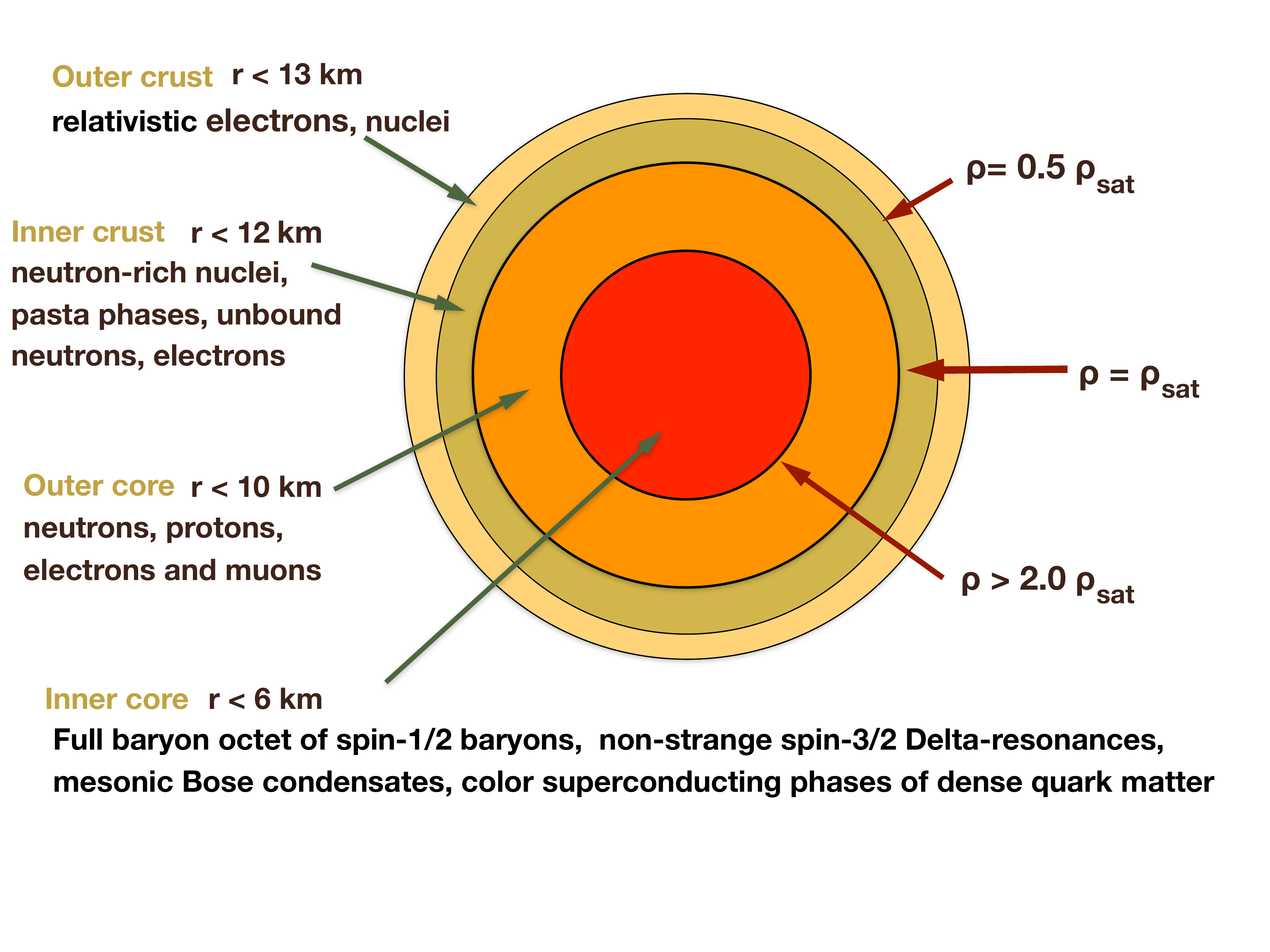}}
\vspace{-1.3cm}
  \caption{Schematic illustration of the interior of a $M = 1.4M_{\odot}$ mass neutron star. The transition density between differen regions (in units of nuclear saturation density $\rho_{\rm sat}$) and the corresponding radial coordinate are indicated. The particle content of each region are indicated as well. Note the possibility of mixed phases involving deconfined quark matter and hypernuclear matter. }
\label{fig:1.1}
\end{figure}

One of the key theoretical challenges of the description of compact stars is the diversity of the possible phases at high densities. A compact star has a (conservatively estimated) mass in the range $1.1\lesssim M/M_{\odot}\lesssim 2.3$, where $M_{\odot}$ is the solar mass, the central densities may reach up to about 10 times the nuclear saturation density ($\rho_{\rm sat} = 0.16$ fm$^{-3}$), and a radius in the range $12 \lesssim R\lesssim 14$~km. A compact star consists roughly of five major regions: the atmosphere, the outer and inner crust, and the outer and inner core, see Fig.~\ref{fig:1.1}. The outer and inner crusts are characterized by the presence of the nuclear clusters immersed in electron gas and a neutron sea in the inner crust. The outer core consists of neutrons, protons, and leptons (mainly electrons with some admixture of muons).  The composition of the inner core starting at about twice the nuclear saturation density is not well established: the multitude of possibilities include hyperonization, the deconfinement phase transition to quark matter, the onset of meson condensation, etc.
For a textbook discussion of the phases of compact stars see, for example, ~\citet{Shapiro:1983du,Glendenning2012compact,weber1999pulsars}.
%--------------------------------------------------------
% --------------------------------------------------------

During the last decade, there have been several breakthrough observational advances that reshaped our understanding of compact stars. These include the measurements of heavy pulsar masses in binaries of neutron stars with white dwarfs, the observation of gravitational waves from double neutron stars mergers, and the simultaneous mass and radius measurement of nearby X-ray emitting neutron stars.  These advances provided crucial insights for the development of theoretical models of dense matter in recent years. Most importantly, they allowed us to narrow down substantially the theoretical range of parameters of the models of dense matter.

This review aims to present recent advances in the studies of the equation of state, composition, and physical manifestations of compact stars that contain hypernuclear cores. This will be done in the framework of the covariant density functional (hereafter CDF) methods~\cite{Serot1997} for the description of dense hypernuclear matter - a method that provides a flexible enough approach to accommodate the information coming from astrophysics and laboratory on hypernuclei. The Lagrangian-based relativistic CDFs of nuclear systems (also known as nuclear relativistic mean-field models) operate with effective degrees of freedom - baryons and meson. They provide a well-motivated and straightforward way to obtain the energy density of matter. At the same time, there is no true fundamental field theory (in the sense of electroweak theory or QCD) associated with these models, i.e., the microscopic underpinning of these models is unknown. Therefore, they are best viewed as CDFs with parameters to be determined from available data in the sense of Kohn-Sham density functional theories which have been successfully applied in many complex many-body systems in various fields, including strongly correlated electronic systems, quantum chemistry, atomic and molecular systems, etc.  The relativistic mean-field model based CDFs have been applied to hypernuclear systems already in 80s and 90s~\cite{Glendenning1985,Glendenning:1991ic,Glendenning:1991es,Huber1998,Huber:1994zz,Schaffner:1995th,Papazoglou:1997uw}.

The motivation to study hypernuclear stars and the interest in hypernuclear CDFs resurged after the observations of two solar mass pulsars in a binary orbit with a white dwarf in 2010~\cite{Demorest2010Nature} when, for the first time since the discovery of a pulsar in 1967, radioastronomy provided us with a significant constraint on the mass of a pulsar. The reason is that hyperons become energetically favorable once the Fermi energy of neutrons exceeds their rest mass. The onset of hyperons reduces the degeneracy pressure of a cold hypernuclear matter, therefore, its equation of state (hereafter EoS) becomes softer than that of nucleonic matter. As a result, the maximum possible mass of a compact star with hyperons decreases to values below $2M_{\odot}$~\cite{BaldoPhysRevC,SchulzePhysRevC}, which is in direct contradiction with the observations.  This contradiction is known as the ``hyperon (or hyperonization) puzzle''.
As discussed below the modern density functionals for the hypernuclear  matter not only resolve the hyperon puzzle but also account for the data coming from gravitational wave physics and X-ray observations of nearby pulsars~\cite{Bednarek:2011gd,Bonanno2012,Weissenborn2012a,Tsubakihara:2012ic,Jiang:2012hy,Massot:2012pf,Providencia:2012rx,Colucci2013,Dalen2014,Gusakov:2014ota,Gomes:2014aka,Banik:2014qja,LopesPhysRevC2014,Maslov2015PhLB,Oertel2015,Drago:2015cea,Maslov:2015wba,Miyatsu:2015kwa,Tolos2016,Oertel:2016xsn,Torres:2016ydl,Fortin2017,Lijj2018a,Lijj2018b,Gomes:2018eiv,Fortin:2020qin}.  The possibility of nucleation of $\Delta$-resonances has been considered along with the hyperonization based on essentially the same arguments which make heavy baryons preferable to highly energetic neutrons~\cite{Sawyer1972ApJ,Waldhauser1987,Waldhauser1988,Weber1989JPG,Choudhury1993,Schurhoff2010}. Again, after the discovery of massive pulsars, the CDF methods have been invoked to treat $\Delta$-resonance admixed (hyper)nuclear matter~\cite{Drago:2014oja,Caibj2015,ZhuPhysRevC2016,Kolomeitsev2017,Lijj2018b,Sahoo:2018xeu,Ribes_2019,Spinella2020:WSBook,Raduta2020,Thapa:2021kfo,Thapa:2020ohp,Dexheimer:2021sxs}.

Because of space limitations we will focused in this review on the CDF approach and did not cover the alternative many-body approaches as applied to hypernuclei and compact stars, which include lattice studies~\cite{Nemura:2007cj, Inoue2019AIPC,Sasaki:2019qnh}, many-body schemes based on G-matrix theory~\cite{Yamamoto:2014jga,Bombaci:2016xzl, Haidenbauer2017EPJA}, Monte-Carlo~\cite{Lonardoni:2012rn,Lonardoni:2013rm,Lonardoni:2014bwa,Gandolfi:2015rvc} and variational~\cite{Togashi:2016fky,Shahrbaf:2019wex,Shahrbaf:2020uau} methods, see also the reviews~\cite{Chatterjee:2015pua,Vidana:2018bdi,Blaschke:2018mqw,Providencia2019}. An overview of the many-body methods with an emphasis on the compact stars can be found, for example,  in ~\citet{Sedrakian2007PrPNP,Oertel2017}.

Compact star properties will be discussed below exclusively with Einstein's theory of general relativity. However, there has been substantial work in recent years on the interpretation of the astrophysics of compact stars within alternative theories of gravity, see,  for example,  \citet{Blazquez-Salcedo:2015ets,Motahar:2017blm,Astashenok:2014pua,Astashenok:2014gda,Astashenok:2014nua} and references therein.

This chapter is organized as follows.  Section~\ref{sec:Astro_constraints} discusses the astrophysical constraints on compact star properties that became available in recent years. The key ideas of the CDF theory are presented in Sec.~\ref{sec:Hyper_DFT}.  Section~\ref{sec:HNS_properties} is devoted to the properties of hypernuclear stars, including the equation of state, composition, and global properties. The recent results on rapidly rotating hypernuclear stars are discussed in Sec.~\ref{sec:Rapid_rotation}.  The interplay between clustering and heavy-baryon degrees of freedom in the warm and low-density nuclear matter is exposed in Sec.~\ref{sec:Clusters}.  The cooling of hypernuclear stars, the finite-temperature equation of state, and universal relations are briefly touched upon in Sections \ref{sec:Cooling} and \ref{sec:Universality}, which is followed by concluding remarks in Sec.~\ref{sec:Conclusions}.

\section{Astrophysical constraints on neutron stars}
\label{sec:Astro_constraints}

Pulsar timing is among the methods that provide information on the masses of pulsars via
measurements of the Keplerian parameters and the observed spin properties of pulsars within neutron star--neutron star and neutron star--white dwarf binaries. The Shapiro delay method of measurement of pulsar mass is based on the observation that electromagnetic radiation experiences a time delay as it passes in the vicinity companion compact object (neutron star or white dwarf) due to its gravitational field. The Shapiro delay method~\cite{Shapiro1964PhRvL} has been successfully applied in binary systems involving millisecond pulsars, where the periodic pulsed signal from the pulsar covers tracks of different length within the space-time continuum depending on whether the pulsar passes in front or behind its binary companion relative to a distant observer. In the general theory of relativity, the Shapiro time delay depends on the companion mass and the degree of inclination of the binary system.  The first high-mass pulsar measurement was carried out for PSR J1614-2230 which is a 3.2-ms pulsar in an 8.7-day orbit with a massive white-dwarf companion in a highly inclined orbit~\cite{Demorest2010Nature}.  The second massive pulsar with high-precision mass measurement is J0348+0432 which is 39 ms pulsar in a 2.46-hour orbit with a white dwarf. In this case, optical observation and modeling of companion white dwarf was used in addition to pulsar timing measurements of Keplerian parameters of binary to place the mass of this millisecond pulsar at $ 2.01\pm 0.04M_{\odot}$~\cite{Antoniadis2013Sci}.  Improved measurements by the NANOGrav place the mass of this pulsar as $1.928(17) M_{\odot}$~\cite{Fonseca2016ApJ}.  The third, most massive (measured with high-precision) neutron star to date is PSR J0740+6620~\cite{Cromartie2020NatAs}, which is a 2.89-ms pulsar in a 4.77-day orbit with a white dwarf. The timing analysis which measures Shapiro delay places the mass of this pulsar at $2.08\pm{0.07}M_{\odot}$ at $68.3\%$ credibility~\cite{Fonseca2021}.  Thus, the timing observations of  three millisecond pulsars J1614-2230, J0348+0432, and J0740+6620 indicate that compact stars with masses as large as $2 M_{\odot}$ exist in Nature.  On the other hand, general relativity predicts that stable compact star sequences end at a maximal mass independent of the equation of state employed.  (The stable configurations are determined by the Bardeen-Thorne-Meltzer criterion \cite{Bardeen1966ApJ}, which implies that a star is stable only as long as its mass is increasing with the central density).  Therefore, we may conclude that this maximum mass predicted by any viable EoS must be at least as large as those observed in the timing observations. In other words, the millisecond pulsar observations place a {\it lower limit} on the maximum mass of a compact star.

With the advent of gravitational wave astronomy and the first measurement by the LIGO and Virgo
collaboration (hereafter LVC) of the gravitational waves from a binary neutron star (hereafter BNS) merger in the GW170817 event ~\cite{Abbott2017a} it became possible to constrain the properties of compact stars through the study of their tidal response.  A weaker signal was detected later in the GW190425 event, which is likely to be a BNS coalescence~\cite{Abbott2020ApJ}.  These measurements, which were carried by the LVC second-generation ground-based gravitational observatories, are based on the idea that a neutron star is deformed in the tidal gravitational field of the companion. To the lowest order, such deformations are described through the star's induced quadrupole moment.  The gravitational wave signal emitted before the merger of a binary carries direct information on the tidal properties of compact stars.  The {\it tidal deformability} $\lambda$ is defined as the coefficient which relates the induced quadrupole $Q_{ij}$ to the perturbing tidal field ${\cal E}_{ij}$ that acts on a star perturbing its shape $Q_{ij} = -\lambda {\cal E}_{ij}$, where $i$ and $j$ label the space coordinates. It is related to the gravitational Love number $k_2$ via the relation
%-------------------------------------------
\begin{equation}
\quad \lambda = \frac{2}{3} k_2 R^5,
\end{equation}
%-------------------------------------------
which exhibits its sensitivity to the radius of the star $R$.  Both $k_2$ and $R$ depend on the EoS  of matter~\cite{Baiotti:2019sew}.  Frequently one uses dimensionless tidal deformability defined as
%-------------------------------------------
\begin{equation}
\label{eq:Lambda}
 \Lambda = \frac{\lambda}{M^5} = \frac{2}{3}k_2 C,
\end{equation}
% -------------------------------------------
where $C= M/R$ is the compactness of the star. Frequently one also defines an effective tidal deformability
as~\cite{FlanaganPhysRevD}
% -------------------------------------------
\begin{equation}
 \tilde \Lambda = \frac{16}{13} \frac{(M_1 + 12M_2) M_1^4\Lambda_1 + (M_2 + 12M_1)M_2^4\Lambda_2}{(M_1 + M_2)^5},
\end{equation}
% -------------------------------------------
involving the masses and tidal deformabilities of both stars.  The analysis of the GW170817 and GW190425 events were carried out for high- and low-spin priors. We give below some characteristic numbers only for low-spin priors as suggested by galactic observations.  The total binary mass values $2.73^{+0.04}_{-0.01}M_{\odot}$ and $3.4^{+0.3}_{-0.1}M_{\odot}$ were deduced from the GW170817 and GW190425 events.  The masses of the components are in the range $1.16 -1.6M_{\odot}$ for GW170817 and $1.46 -1.87M_{\odot}$ for GW190425 event.  The analysis of GW170817 event led to an upper limit $ \Lambda_{1.4} < 700$ for a $1.4M_{\odot}$ star ($90\%$ confidence). In the case of the GW190425 event, $\tilde \Lambda\le 600$ has been inferred for the binary mass range indicated above.

Neutron star surfaces are emitting $X$-rays due to their thermal heating by currents associated with the particle flows in the magnetosphere. The location of the emitting hot spots reflects the structure and topology of the magnetosphere.  Due to the rotation of the star, the hot spots are generating pulsed emission. The thermal $X$-ray emission pulse profile of millisecond pulsar J0030+0451 has been used by the NICER team to place limits on its mass {\it and } radius~\cite{Miller2019ApJ,Riley2019ApJ}. The procedure involves modeling the soft $X$-ray pulses produced by the rotation of hot spots on the surface of the star and fitting them to the NICER waveform data. In doing so it was assumed that the star's emitting atmosphere is ionized hydrogen and that the magnetic field does not play any role. The two independent analyses predict ($68\%$ credible interval) $M = 1.44^{+0.15}_{-0.14} M_{\odot}$, $R = 13.02^{+1.24}_{^-1.06}$~km ~\cite{Miller2019ApJ} and $M=1.34^{+0.15}_{-0.16} M_{\odot}$, $R= 12.71^{+1.14}_{-1.19}$~km~\cite{Riley2019ApJ}.  The same collaboration has also measured the radius of PSR J0740+662 to be $R= 13.7^{+2.6}_{-1.5}$~km~\cite{Miller2021} and $R= 12.39^{+1.3}_{-0.98}$~km~\cite{Riley:2021pdl} ($68\%$ credible interval).  In addition to the above constraints, the masses of neutron stars in binaries have been measured with high accuracy which lie in the range $1.2\lesssim M/M_{\odot}\lesssim 1.6$ with a significant concentration around the value $1.4~M_{\odot}$~\cite{Ozel:2016oaf,Lattimer:2019eez}. Furthermore, the moment of inertia of a neutron star is expected to be measured in the double pulsar system PSR J0737-3039~\cite{Lattimer2005ApJ}, where both masses of stars are already accurately determined by observations.

\section{Density functionals for hypernuclear matter}
\label{sec:Hyper_DFT}

Density functional theory offers a flexible framework to study the equilibrium thermodynamics of nuclear and neutron star matter. In nuclear physics, a particular class of density functionals can be obtained within the so-called {\it ``relativistic mean-field''} models of nuclear matter, which by construction posses the main feature of density functional theory: the potential energy of the zero-temperature system is a functional of (energy) density alone. The self-energies in these theories are evaluated either in the Hartree or Hartree-Fock theories of many-body theory, the first approximation being the simplest realization of such scheme. The Hartree-Fock theories allow one to introduce the pion contribution explicitly in the density functional, which could be an advantage for a detailed treatment of the tensor force. Once the density functional is constructed the parameters of the relativistic Lagrangian are adjusted to reproduce the laboratory data within a range that is compatible with other constraints, such as those available from compact star observations. The microscopic {\it ab initio} many-body calculations are treated as data, i.e., they constrain the admissible range of parameters entering the functional.
(In passing we note that density functionals have been derived directly from microscopic theories, but this will not be reviewed here). An advantage of the approach taken here is the straightforward extension of the density functional from nuclear to hypernuclear and $\Delta$-resonance matter. The fast numerical implementations allow us to scan a large parameter space associated with the density functional, which can become quite large as one includes the heavy baryons in the functional.

\subsection{Relativistic density functional with density-dependent couplings}
\label{sec:Hyper_DDME2}

\begin{figure}[t]
  \begin{center}
  \includegraphics[width=12cm]{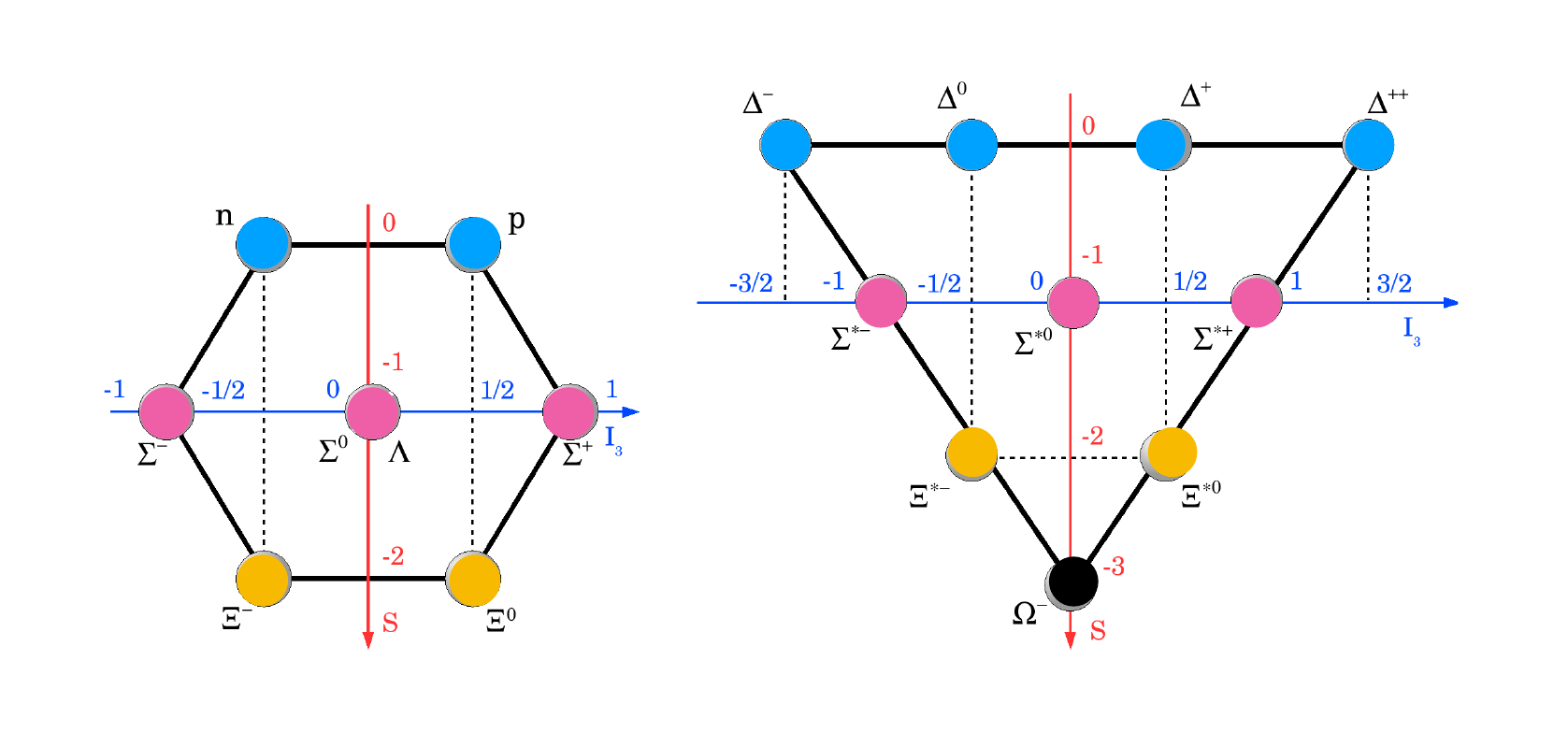}
\end{center}
\vskip -0.5cm
\caption{An illustration of the spin-1/2 octet of baryons (left) and spin-3/2 decuplet of resonances (right), where the vertical axis is the strangeness and the horizontal axis is the third component of the
isospin.}
\label{fig:1.2}
\end{figure}

In this section, we briefly review the construction of the density functional of hypernuclear matter starting from a relativistic Lagrangian and adopting the Hartree approximation. The Hartree-Fock scheme is discussed in detail elsewhere~\cite{Lijj2018a}. We will use a particular class of such functionals that assign a density dependence to the coupling constants describing the meson-baryon interactions, which incorporate modifications of the interaction due to changes in the density of the medium in which baryons and mesons are
embedded~\cite{Typel1999,Lalazissis2005,Typelparticles2018}. The scheme discussed below will include successively the $J=1/2$ baryon octet and $J=3/2$ resonances; see Fig.~\ref{fig:1.2} for an illustration of the octet and decuplet arrangement of the heavy baryons and their quantum numbers.
%--------------------------------------------------------
% --------------------------------------------------------
The Lagrangian of stellar matter with baryonic degrees of freedom can be written as
\begin{equation}
  \label{eq:Lagrangian}
 \mathscr{L} = \mathscr{L}_b +  \mathscr{L}_m +  \mathscr{L}_l +  \mathscr{L}_{\rm em},
\end{equation}
% --------------------------------------------------------
where the baryon Lagrangian is given by
% --------------------------------------------------------
\begin{eqnarray}
  \label{eq:lagrangian} \nonumber
  \mathscr{L}_b & = &   \sum_b\bar\psi_b\bigg[\gamma^\mu \left(i\partial_\mu-g_{\omega b}\omega_\mu
  - \frac{1}{2} g_{\rho B}\vectau\cdot\vecrho_\mu\right)
                   - (m_b - g_{\sigma b}\sigma)\bigg]\psi_b ,\\
\end{eqnarray}
% --------------------------------------------------------
where the $b$-sum is over the $J^P = \frac{1}{2}^+$ baryon octet, $\psi_b$ are the baryonic Dirac fields of baryons with masses $m_b$, and $\sigma,\omega_\mu$, and $\vecrho_\mu$ are the mesonic fields which mediate the interaction among baryon fields. The coupling constants $g_{mb}$, in general, are density-dependent.
The mesonic part of the Lagrangian is given by
% --------------------------------------------------------
\begin{eqnarray}
\mathscr{L}_m &=& \frac{1}{2}
\partial^\mu\sigma\partial_\mu\sigma-\frac{m_\sigma^2}{2} \sigma^2 -
\frac{1}{4}\omega^{\mu\nu}\omega_{\mu\nu} + \frac{m_\omega^2}{2}
               \omega^\mu\omega_\mu - \frac{1}{4}\vecrho^{\mu\nu}\cdot \vecrho_{\mu\nu}
               + \frac{m_\rho^2}{2} \vecrho^\mu\cdot\vecrho_\mu ,\nonumber\\
\end{eqnarray}
% --------------------------------------------------------
where $m_{\sigma}$, $m_{\omega}$, and $m_{\rho}$ are the meson masses
and $\omega_{\mu\nu}$ and $\vecrho_{\mu\nu}$ represent the strength tensors of vector mesons
% --------------------------------------------------------
\begin{eqnarray}             
\omega_{\mu \nu} & = \partial_{\mu}\omega_{\nu} - \partial_{\mu}\omega_{\nu} ,\qquad
\boldsymbol{\rho}_{\mu \nu} & = \partial_{\nu}
\boldsymbol{\rho}_{\mu} - \partial_{\mu}\boldsymbol{\rho}_{\nu}.
\end{eqnarray}
% --------------------------------------------------------
 The leptonic Lagrangian is given by
% --------------------------------------------------------
\begin{eqnarray}             
\mathscr{L}_l &=& \sum_{\lambda}\bar\psi_\lambda(i\gamma^\mu\partial_\mu -
      m_\lambda)\psi_\lambda,
\end{eqnarray}
% --------------------------------------------------------
where $\psi_\lambda$ are lepton fields (in stellar matter the $\lambda$-summation includes electron, muons, and
at high temperatures the three flavors of neutrinos when neutrinos are trapped) and $m_\lambda$ are their masses. Finally, if the stellar matter is permeated by sizable magnetic fields (which is the case, for example, in magnetars) then we need to include the electromagnetism via its gauge-field Lagrangian
% --------------------------------------------------------
\begin{equation}
 \mathscr{L}_{\rm em} = - \frac{1}{4}F^{\mu\nu}F_{\mu\nu},
\end{equation}
% --------------------------------------------------------
where $F^{\mu\nu}$ is the electromagnetic field strength tensor and replace the partial derivatives by their gauge-invariant counterparts $D_{\mu}= \partial_{\mu} + ie Q A_{\mu}$ where $A_\mu$ is the electromagnetic vector potential and $eQ$ is the charge of the particle~\cite{Thapa:2020ohp,Dexheimer:2021sxs}.

The baryonic Lagrangian can be extended to include the non-strange $J=3/2$ members of the baryons decuplet which is the quartet of $\Delta$-resonances, see Fig.~\ref{fig:1.2}, by adding to the Lagrangian \eqref{eq:Lagrangian} the term
% --------------------------------------------------------
\begin{eqnarray}
\mathscr{L}_d = \sum_{d}\bar\psi^{\nu}_{d}\bigg[\gamma^\mu \left(i\partial_\mu-g_{\omega d}\omega_\mu
  - \frac{1}{2} g_{\rho d}\vectau\cdot\vecrho_\mu\right)
                   - (m_d- g_{\sigma d}\sigma)\bigg]\psi_{d\nu} ,\nonumber\\
\end{eqnarray}
% --------------------------------------------------------
where the $d$-summation is over the resonances described by the Rarita-Schwinger fields $\psi_{d\nu}$. Furthermore, to describe the interactions between the strange particles the Lagrangian \eqref{eq:Lagrangian} can be extended to include the interactions in the hypernuclear sector via two additional (hidden strangeness) mesons $\sigma^*$ and $\phi$
%-------------------------------------------------------------------------------
\begin{eqnarray}
\mathscr{L}_{m'} = \frac{1}{2}
\partial^\mu\sigma^*\partial_\mu\sigma^*-\frac{m_\sigma^{*2}}{2} \sigma^{*2}
- \frac{1}{4}\phi_{\mu\nu}\phi^{\mu\nu} + \frac{1}{2}m_{\phi}^2\phi_{\mu}\phi^{\mu},
\end{eqnarray}
%-------------------------------------------------------------------------------
with field strength tensor $\phi_{\mu \nu}  = \partial_{\nu}\phi_{\mu} - \partial_{\mu}\phi_{\nu}$.
These mesons do not couple to nucleons, i.e., $g_{\sigma^* N}= g_{\phi N}=0$, see however
Eq.~\eqref{eq:phiome relation} below.

Let us now turn to the coupling constants in the nucleonic sector. These are given by their values at the saturation density $\rho_{\rm sat}$ and by functional form describing their dependence on baryon density $\rho_B$
%-------------------------------------------------------------------------------
\begin{equation}
 g_{iN}(\rho_B) = g_{iN}(\rho_{\rm sat})h_i(x),
\end{equation}
% -------------------------------------------------------------------------------
where $x = \rho_B/\rho_{\rm sat}$ and 
% -------------------------------------------------------------------------------
\begin{eqnarray} \label{eq:h_functions}
  h_i(x) =\frac{a_i+b_i(x+d_i)^2}{a_i+c_i(x+d_i)^2},~i=\sigma,\omega,\quad
h_\rho(x) = e^{-a_\rho(x-1)}.\nonumber
\end{eqnarray}
%-------------------------------------------------------------------------------
The density dependence of the couplings implicitly takes into account many-body correlations that modify the interactions in the medium. In the following we will adopt the DDME2 parametrization~\cite{Lalazissis2005}; the values of the parameters are listed for completeness in Table~\ref{tab:1}. 
% -------------------------------------------------------------------------------
\begin{table}[t]
\tbl{ The values of parameters of the DDME2 CDF.}
{
\begin{tabular}{ccccccc}
\hline \hline
Meson ($i$) & $m_i$ (MeV) & $a_{i}$ & $b_{i}$ & $c_{i}$ & $d_{i}$ & $g_{iN}$ \\
\hline
$\sigma$ & 550.1238 & 1.3881 & 1.0943 & 1.7057 & 0.4421 & 10.5396 \\
$\omega$ & 783 & 1.3892 & 0.9240 & 1.4620 & 0.4775 & 13.0189 \\
$\rho$ & 763 & 0.5647 & & & & 7.3672 \\
  \hline
\end{tabular}
}
\label{tab:1}
\end{table}

% -------------------------------------------------------------------------------
The masses of $\omega$ and $\rho$ mesons are taken to be their free values.
The five constraints $h_{i }(1)=1$, $h_i^{\prime\prime}(0)=0$ and $h^{\prime\prime}_{\sigma}(1)=h^{\prime\prime}_{\omega }(1)$ allow one to reduce the number of free parameters in isocalar-scalar and iso-scalar-vector sector to three.  Three additional parameters in this channel are $g_{\sigma N} (\rho_{\rm sat}), g_{\omega N} (\rho_{\rm sat})$ and $m_\sigma$ - the mass of the phenomenological $\sigma$ meson.  With two additional parameter in the isovector-vector channel, the parameterization has in total eight parameters (seven entering the definition of the coupling constants and the mass of the $\sigma$-meson) which are adjusted to reproduce the properties of symmetric and asymmetric nuclear matter, binding energies, charge radii, and neutron skins of spherical nuclei.

The ground-state expectation values of mesons in the mean-field and infinite system approximations are given by
%------------------------------------------------------------------------
\begin{eqnarray}
      &&  m_{\sigma}^2\sigma = \sum_{b} g_{\sigma b}n_{b}^s + \sum_{d} g_{\sigma d}n_{d}^s,
 \quad m_{\sigma^{*}}^2 \sigma^{*} = \sum_{b} g_{\sigma^{*}   b} n_{b}^s,\\
  && m_{\omega}^2\omega_{0}= \sum_{b} g_{\omega b}n_{b} +\sum_{d} g_{\omega d}n_{d},
  \quad m_{\phi}^2\phi_{0}= \sum_{b} g_{\phi b}n_{b} ,\\
&&  m_{\rho}^2\rho_{03}= \sum_{b} g_{\rho b}
  \boldsymbol{\tau}_{b3}n_{b} + \sum_{d} g_{\rho d}  \boldsymbol{\tau}_{d3}n_{d},
\end{eqnarray}
%------------------------------------------------------------------------
where the scalar and baryon (vector) number densities are defined
 for the baryon octet as
$n_{b}^s= \langle\bar{\psi}_b \psi_b\rangle$  and~$n_{b}=                                                       
\langle\bar{\psi}_b \gamma^0 \psi_b\rangle$, respectively. For~
  the $\Delta$-resonances, these are defined as $n_{d}^s=                                                       
  \langle\bar{\psi}_{d\nu} \psi^\nu_d\rangle$ and $n_{d}=                                                       
  \langle\bar{\psi}_{d\nu} \gamma^0 \psi^\nu_d\rangle$,
  respectively. The~effective (Dirac) baryon masses in the same
  approximation are given by
%------------------------------------------------------------------------
\begin{equation}
m_{b}^* = m_b - g_{\sigma b}\sigma - g_{\sigma^*b}\sigma^*, \quad
m_{d}^* = m_d - g_{\sigma d}\sigma.
\end{equation}
%------------------------------------------------------------------------
Given the Lagrangian density (\ref{eq:lagrangian}), the energy stress tensor can be constructed
%------------------------------------------------------------------------
\begin{eqnarray}
T^{\mu\nu} = \frac{\partial \mathscr{L}}{\partial (\partial_\mu\varphi_i)}\partial^\nu \varphi_i -  g^{\mu\nu}\mathscr{L},
\end{eqnarray}
% ------------------------------------------------------------------------
where $\varphi_i$ stands generically for a boson or fermion field. Then,
its diagonal elements define the energy density and pressure
% ------------------------------------------------------------------------
\begin{eqnarray}
{\cal E} = \langle T^{00}\rangle, \quad P = \frac{1}{3}\sum_i\langle T^{ii}\rangle,
\end{eqnarray}
% ------------------------------------------------------------------------
where the brackets refer to statistical averaging. Explicitly one finds
%------------------------------------------------------------------------
\begin{eqnarray}
  P & = & - \frac{m_\sigma^2}{2} \sigma^2 -\frac{m_\sigma^{*2}}{2} \sigma^{*2}
          + \frac{m_\omega^2}{2} \omega_0^2 + \frac{m_\rho^2 }{2} \rho_{03}^2
          +  \frac{m_\phi^2 }{2} \phi_0^2
\nonumber\\
&+ & \frac{1}{3}\sum_{b,d} \frac{2J_{b,d}+1}{2\pi^2}\int_0^{\infty}\!\!\! \frac{dk\ k^4 }{E_k^{b,d}}
\left[f(E_k^{b,d}-\mu_{b,d}^*)+f(E_k^{b,d}+\mu_{b,d}^*)\right]\nonumber\\
& +&\frac{1}{3\pi^2} \sum_{\lambda} \int_0^{\infty}\!\!\!  \frac{dk\ k^4 }{E_k^\lambda}
\left[f(E_k^{\lambda}-\mu_\lambda)+f(E_k^{\lambda}+\mu_\lambda)\right]
\end{eqnarray}
%------------------------------------------------------------------------
and
%------------------------------------------------------------------------
\begin{eqnarray}
  {\cal E} &=& \frac{m_\sigma^2}{2} \sigma^2 +\frac{m_\sigma^{*2}}{2} \sigma^{*2} +
                                \frac{m_\omega^2 }{2} \omega_0^2 + \frac{m_\rho^2}{2} \rho_{03}^2 
                              +  \frac{m_\phi^2 }{2} \phi_0^2
                                \nonumber\\
&+& \sum_{b,d} \frac{2J_{b,d}+1}{2\pi^2}  
\int_0^{\infty}dk \ k^2 E_k^{b,d} \left[f(E_k^{b,d}-\mu_{b,d}^*)
+f(E_k^{b,d}+\mu_{b,d}^*)\right] \nonumber\\
&+& \sum_{\lambda} \int_0^{\infty} \!\!\!\frac{dk}{\pi^2}
k^2E_k^\lambda\left[f(E_k^\lambda-\mu_\lambda)
+f(E_k^\lambda+\mu_\lambda)\right],
\end{eqnarray}
%------------------------------------------------------------------------
where $2J_{b,d}+1$ is the baryon degeneracy factor with $J_{b}=1/2$ for baryon octet
and $J_{d}=3/2$ for $\Delta$-resonances,
%------------------------------------------------------------------------
\begin{eqnarray}
&& \mu_{b} = \mu_b^* + g_{\omega b}\omega_{0} + g_{\phi b}\phi_{0} 
+ g_{\rho b} \tau_{b3} \rho_{03} + \Sigma^{r}, \\
  && \mu_{d} = \mu_d^* + g_{\omega d}\omega_{0} +  g_{\rho d}
    \tau_{d3} \rho_{03} + \Sigma^{r},
\end{eqnarray}
%------------------------------------------------------------------------
are the baryon chemical potentials, $I_3$ is the third component of baryon isospin, $E_k^{b,d} = \sqrt{k^2+m^{*2}_{b,d}}$ and $E_k^{\lambda}=\sqrt{k^2+m_\lambda^2}$ are the single particle energies of baryons and leptons respectively, and $f(E) = [1+\exp(E/T)]^{-1}$ is the Fermi distribution function at temperature $T$.  The lepton mass $m_\lambda$ can be taken equal to its free-space value.  The self-energy $\Sigma^r$ arises due to the density-dependence of the coupling constant and guarantees the thermodynamic consistency of the theory, i.e.,
the fact that the thermodynamic relation
%------------------------------------------------------------------------
\begin{equation}
  P = \rho^2 \frac{\partial }{\partial \rho}\left(\frac{\cal E}{\rho}\right),
\end{equation}
% ------------------------------------------------------------------------
is fulfilled.

\subsection{Hyperonic and $\Delta$-resonance couplings}
\label{sec:Couplings}

The lack of reliable information on the hyperon-nucleon and hyperon-hyperon interactions prevents a high-precision determination of the parameters entering the Lagrangian \eqref{eq:lagrangian}.  The SU(3) flavor symmetric model allows one to determine the magnitude of the coupling constants appearing in the Lagrangian \eqref{eq:lagrangian} based on symmetry argument and the principles of the ``eightfold way'' of elementary particle classification in particle physics~\cite{Swart1963}. The SU(3) symmetry in flavor space is, of course, broken by the strange quark mass at the densities and temperatures relevant to compact stars. Nevertheless, it provides some guidance in instances where no or little information is available.

Within this model, the spin-$1/2$ baryons and mesons are arranged in octets, which is the lowest non-trivial irreducible representation of the symmetry group.  The interaction part of the SU(3) invariant Lagrangian describing the coupling of baryons and mesons is constructed using matrix representations for the baryon $J^P = 1/2^+$ octet of baryons $B$ and $J^P=1^-$ meson octet ($M_8$) which is supplemented by meson singlet ($M_1$) which allows describing physical mesons via mixing mechanism. The Lagrangian contains linear combinations of the antisymmetric ($F$-type), symmetric ($D$-type), and singlet ($S$-type) scalar contributions (using the standard notations)
%---------------------------------------------------------------
\begin{eqnarray}
\mathscr{L}_{SU(3)} &=& -g_8\sqrt{2}[\alpha\text{Tr}([\bar{B},M_8]B) + (1-\alpha)\text{Tr}(\{\bar{B},M_8\}B)]
  \nonumber\\
  &-&\frac{g_1}{\sqrt{3}}\text{Tr}(\bar{B}B)\text{Tr}(M_1),
\end{eqnarray}
%---------------------------------------------------------------
where  $g_8$ and $g_1$ denote the meson octet and singlet coupling constant respectively
and $\alpha=F/(F+D)$ with $0\le \alpha \le 1$.

The physical mesons $\omega$ and $\phi$ then appear as a mixture of the 
$\omega_0$ and $\omega_8$ members of the vector meson nonet: 
% ----------------------------------------------------
\begin{equation}
  \left(  \begin{array}{c}
            \omega_8\\
            \omega_0 
            \end{array}
          \right)  =
            \left(  \begin{array}{cc}
            \cos\theta_V & \sin\theta_V\\
            -\sin\theta_V & \cos\theta_V
            \end{array}\right) \left(  \begin{array}{c}
            \omega\\
            \phi 
            \end{array}
          \right),
        \end{equation}
% ----------------------------------------------------
        where $\theta_V$ is the vector mixing angle. Within this mixing scheme the
         coupling of a baryon to the physical $\omega$-meson is given by 
%----------------------------------------------------
\begin{equation}\label{vector_mixing}
  g_{B\omega} = \cos\theta_V g_1 + \sin\theta_V \frac{g_8}{\sqrt{3}}\delta_B, \qquad B \in \{N, \Xi,\Lambda,\Sigma\}, 
\end{equation}
% ----------------------------------------------------
where $\delta_N = 4\alpha_V-1$, $\delta_\Xi = 1+2\alpha_V$ and $\delta_\Sigma = -\delta_\Lambda =2( 1-\alpha_V)$. It is convenient to express the coupling of hyperons to mesons by using the nucleonic couplings as normalization
% ----------------------------------------------------
\begin{eqnarray}
  R_{\omega B} = \frac{g_{\omega B}}{g_{\omega N}} =
  \frac{1-\frac{g_8}{g_1\sqrt{3}}\delta_B\tan \theta_V}{1-\frac{g_8}{g_1\sqrt{3}}(1-4\alpha_V)\tan \theta_V}.
\end{eqnarray}
% ----------------------------------------------------
The $\phi$-meson couplings can be obtained from those for $\omega$ meson by the substitution
$\cos\theta_V \rightarrow -\sin\theta_V$ and $\sin\theta_V\rightarrow\cos\theta_V$.
In general, it is possible that $\phi$ meson couples to the nucleon with the coupling given by 
%------------------------------------------------
\begin{align}\label{eq:phiome relation}
R_{\phi N}&= -
\frac{\tan \theta_V -\frac{g_8}{g_1\sqrt{3}}\delta_N}{1+\frac{g_8}{g_1\sqrt{3}}\delta_N\tan \theta_V}.
\end{align}
% -------------------------------------------------
The coupling for the isovector $\rho$ meson is given by  
% -------------------------------------------------
\begin{eqnarray}\label{vector_relations} 
  &&g_{\rho N} = g_8, \quad  R_{\rho\Xi} = - (1-2\alpha_V), \quad 
     R_{\rho\Sigma} = 2  \alpha_V, \quad R_{\rho\Lambda} = 0.
\end{eqnarray}
% -------------------------------------------------
Note that the $\rho$ couplings vanish exponentially at high densities according to Eq.~\eqref{eq:h_functions}
and their effect on the properties of dense matter (beyond the threshold for the onset of hyperons)
is small.

The approximate equality of the masses of $\omega$ and $\rho$-mesons implies that
the mixing is ideal, in which case 
 $\phi$ meson is a pure $\bar{s}s$ state and the mixing angle is given by the 
\textit{ideal mixing} value $\tan \theta^{*}_V = 1/\sqrt{2}.$
Since the nucleon does not couple to pure strange meson $\phi$,
Eq.~\eqref{eq:phiome relation} implies that this is the case when 
$g_1 = \sqrt{6}\,g_8$ and  $\alpha_V = 1$, the latter being the universality assumption for the (electric) $F/(F + D)$ ratio, i.e., only $F$-type coupling is non-zero. In this case the coupling constants
of heavy baryons are related to those of the nucleon as in the additive quark model.

The couplings in the case of  the scalar mesons $\sigma$ and $\sigma^\ast$,
are obtained from those of $\omega$ and $\phi$, respectively, with the replacements $\omega\to \sigma$, $\phi\to \sigma^\ast$, and changing the vector indices to scalar ones $V\to S$.
In the case of $\sigma$-meson the coupling constants are given by 
% ----------------------------------------------------
\begin{equation}
  g_{B\sigma}    =  \cos\theta_S~g_1 +  \delta_{B}g_8 \sin\theta_S/\sqrt{3},\\
\end{equation}
where in the definitions of $\delta_B$ the scalar ratio $\alpha_S$ appears instead of
its vector counterpart.  It can be shown that the coupling defined in this manner
obey the following relation~\cite{Colucci2013}
% ----------------------------------------------------
\begin{eqnarray}
  \label{coupling_variation}
2(g_{N\sigma} + g_{\Xi\sigma}) = 3g_{\Lambda\sigma} +
g_{\Sigma\sigma},
  \end{eqnarray}
% ----------------------------------------------------
  which is valid for arbitrary values of the four parameters $\alpha_S$, $g_1$, $g_8$ and $\theta_S$.
  It is easy to verify that Eq.~\eqref{coupling_variation} is satisfied for the coupling constants in the SU(6) spin-flavor symmetric quark model. Table~\ref{tab:2} lists the couplings within this model.
% -------------------------------------------------------------------
\begin{table}[t]
\tbl{The ratios of the couplings of hyperons in the SU(6) spin-flavor model.}
{
\begin{tabular}{cccccc}
    \hline
 $Y\backslash R$ & $R_{\sigma Y}$  & $R_{\sigma^\ast Y}$      & $R_{\omega Y}$             &$R_{\phi Y}$                  &$R_{\rho Y}  $             \\
  \hline
$\Lambda $ & 2/3  &  $-\sqrt{2}/3$   &   2/3 &  $\sqrt{2}/3$  & 0 \\
$\Sigma $ & 2/3  &  $-\sqrt{2}/3$   &   2/3   &  -$\sqrt{2}/3$  &2   \\
$\Xi$    &    1/3  &  $-2\sqrt{2}/3$ &    1/3 &  $-2\sqrt{2}/3$  &    1 \\
\hline
  \hline
\end{tabular}
}
\label{tab:2}
\end{table}
%--------------------------------------------------------

The information on the couplings of hyperons to the scalar mesons can
be obtained from the fits to their potentials in nuclear matter and nuclei within a particular model.
%--------------------------------------------------------
\begin{figure}[t]
 \begin{center}
   \includegraphics[width=9.5cm]{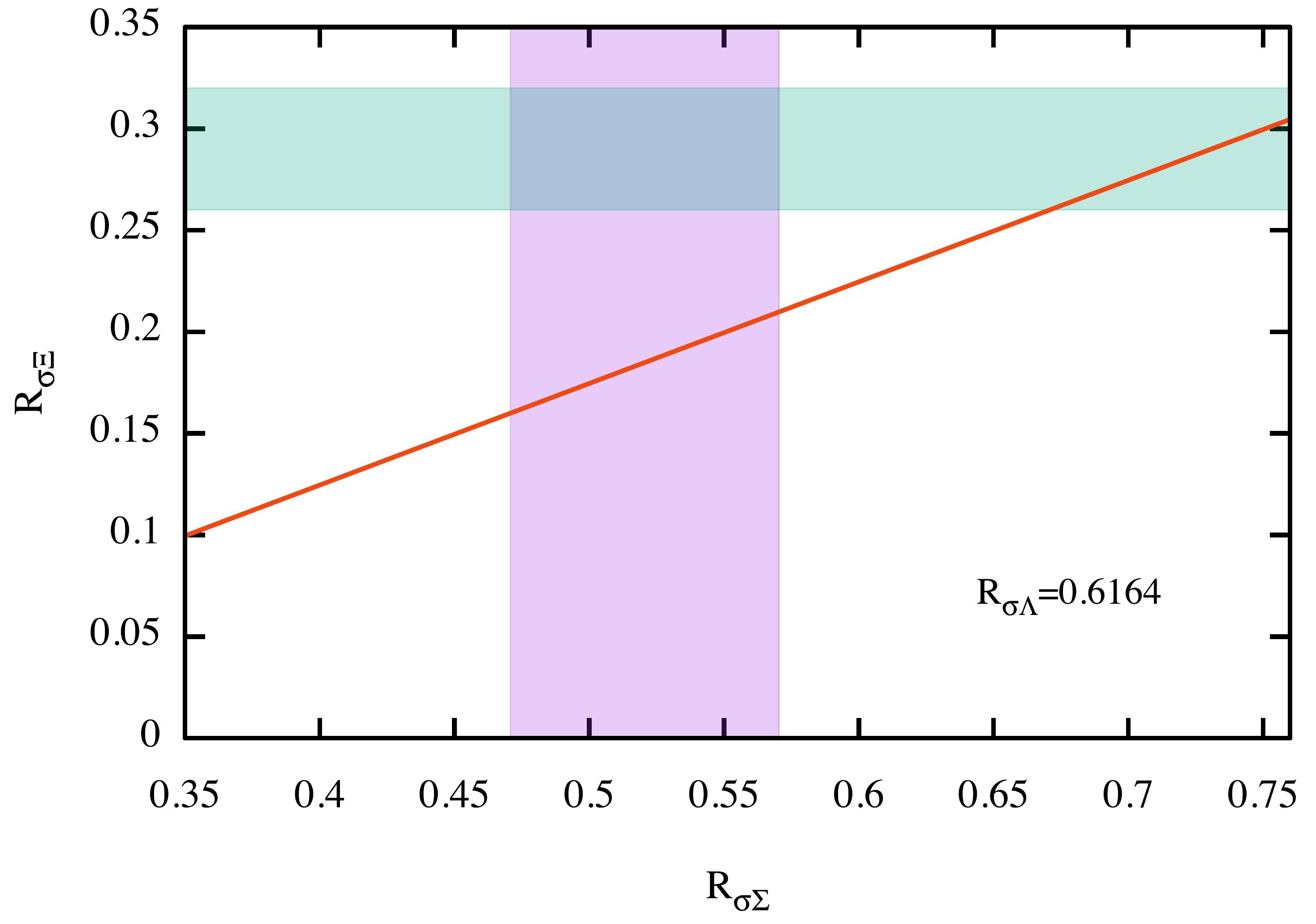} \end{center}
 \vskip -0.5cm
\caption{The ranges of coupling constants $R_{\sigma\Sigma}$ and
  $R_{\sigma\Xi}$ (shaded areas) compatible with
  experimental/theoretical information and relation
  \eqref{coupling_variation} (solid line) for DDME2 CDF and fixed
  $R_{\sigma\Lambda}= 0.6164$. The combined area of $R_{\sigma\Sigma}$
  and $R_{\sigma\Xi}$ has no overlap with prediction based on the
  SU(3) flavor symmetry, indicating its breaking in the scalar-meson
  sector.}
\label{fig:1.3}
\end{figure}
% --------------------------------------------------------
For example, the single $\Lambda$-hypernuclei has been used to fix the
value of $g_{\Lambda\sigma}$ coupling ~\cite{Dalen2014}.  Similarly,
the coupling to the $\sigma^*$ is obtained from the fits to the
binding of double $\Lambda$ hypernuclei
\cite{Fortin2017}. Furthermore, density functionals have been adapted
to treat multi-strange nuclei
hypernuclei~\cite{Khan:2015bxa,Margueron:2017eeq,Guven:2018sgo}.

  Figure~\ref{fig:1.3} shows the range of the (dimensionless)
  couplings $R_{\sigma\Sigma}$ and $R_{\sigma\Xi}$ which covers the
  potential depths ranges $U_{\Sigma}(\rho_{\rm sat}) = [-10:
    +30]$~MeV and $U_{\Xi}(\rho_{\rm sat}) = [-24:0]$~MeV at nuclear
  saturation density.  The value $U_{\Xi}(\rho_{\rm sat}) = -24$~MeV
  has been given in \cite{Friedman2021} and is much deeper than the
  one expected from Lattice 2019
  results~\cite{Inoue2019AIPC,Sasaki:2019qnh}. The fixed value of the
  $R_{\sigma\Lambda}= 0.6164$~\cite{Dalen2014} has been used.  It is
  seen that in contrast to the SU(6) model, the values of
  $R_{\sigma\Xi}$ predicted by relation \eqref{coupling_variation} do
  not intercept the overlap area of $R_{\sigma\Sigma}$ and
  $R_{\sigma\Xi}$ couplings, which indicates the breaking of
  corresponding symmetry.

 The results pertaining hyperonic compact stars
 below have been obtained with the parameters $R_{\sigma\Lambda} = 0.6106$,
 $R_{\sigma\Sigma} = 0.4426$, and $R_{\sigma\Xi} = 0.3024$~\cite{Lijj2018b}
 in the framework of the DDME2 CDF extension to hypernuclear sector. 

Most studies of $\Delta$s in dense matter have been carried out in the 
 relativistic mean-field  approach
\cite{Waldhauser1987,Waldhauser1988,Weber1989JPG,Choudhury1993,Caibj2015,Lavagno2010}, the density-dependent CDF approach~\cite{Drago:2014oja,Kolomeitsev2017,Lijj2018b}, the relativistic Hartree-Fock approach \cite{Weber1989NPA,ZhuPhysRevC2016} and the quark-meson coupling model~\cite{Motta2020delta}. 
 The interactions of $\Delta$-resonance within the matter are not well known.
Some information on the $\Delta$-potential in the isospin symmetric nuclear matter is available from the analysis of scattering of electrons and pions off nuclei with $\Delta$-excitation~\cite{Koch:1997ei,Connell1990PhRvC,Wehrberger1989NuPhA,Horikawa1980NuPhA,Nakamura2010PhRvC} and photo-absorption~\cite{Alberico1994PhLB,Riek2009}. The extracted value of the potential is~\cite{Drago:2014oja}
%---------------------------------------------------------------
\begin{equation}
  -30~\textrm{MeV} +V_{N} (\rho_{\rm sat})\le  V_\Delta(\rho_{\rm sat})\le  V_N(\rho_{\rm sat}),
\end{equation}
% ---------------------------------------------------------------
where $V_{N} (\rho_{\rm sat})$ is the nucleon isoscalar potential at the saturation density.
The $\Delta$-resonance production in heavy-ion collisions is another channel of information,
where however collective dynamics of nuclear matter comes into play~\cite{Cozma2016PhL,Cozma2021,Ono2019PhRvC,Xu2019PrPNP}. Numerical simulations provide hints towards the values of the potential in the range~\cite{Kolomeitsev2017}
%---------------------------------------------------------------
\begin{equation}
  V_{N} (\rho_{\rm sat})\le  V_\Delta(\rho_{\rm sat})\le  2/3V_N(\rho_{\rm sat}).
\end{equation}
% ---------------------------------------------------------------
The isovector meson-$\Delta$-resonance couplings are not known. In the following
we use the ratios $R_{m\Delta } = g_{m\Delta}/g_{mN}$ to describe the $\Delta$-resonance
couplings. In the recent work, these parameters have been varied in the range~\cite{Lijj2018b,Li2019PhRvC}
%---------------------------------------------------------------
\begin{equation}
R_{\rho\Delta} = 1,  \quad  0.8 \le R_{\omega\Delta} \le 1.6,     \quad            
R_{\sigma\Delta} = R_{\omega\Delta} \pm 0.2,
\end{equation}
% ---------------------------------------------------------------
to explore the consequences of the inclusion of $\Delta$-resonance in the CDF of (hyper)nuclear matter.

\subsection{Characteristics of nuclear matter close to saturation}
\label{sec:Characteristics}

As is well known, the EoS of nuclear matter can be parametrized via
an expansion in the vicinity of saturation density and isospin-symmetrical limit via a double-expansion in the Taylor series:
%-------------------------------------------------------                                                                                            
\begin{eqnarray}
\label{eq:Taylor_expansion}
  E(\chi, \delta) & \simeq & E_{\text{sat}} + \frac{1}{2!}K_{\text{sat}}\chi^2
                             + \frac{1}{3!}Q_{\text{sat}}\chi^3  + E_{\text{sym}}\delta^2 + L_{\text{sym}}\delta^2\chi
                  + {\mathcal O}(\chi^4,\chi^2\delta^2),\nonumber\\
\end{eqnarray}
% ---------------------------------------------------------
where $\chi=(\rho-\rho_{\text{sat}})/3\rho_{\text{sat}}$, $\delta = (\rho_{n}-\rho_{p})/\rho$ with $\rho_n$ and $\rho_p$ being the neutron and proton densities. The coefficients of the expansion are known as {\it incompressibility} $K_{\text{sat}}$, the {\it skewness} $Q_{\text{sat}}$, the {\it symmetry energy} $E_{\text{sym}}$ and the {\it slope parameter} $L_{\text{sym}}$.  The definitions of parameters are standard, see~\citet{Margueron2018a}.  These four low-order coefficients can be constrained from experimental data on nuclear systems.  The higher-order terms are less constrained~\cite{Margueron2018a,Zhangnb2019}.

 While the CDFs provide full access to the spectra of particles, the matter compositions, etc. the expansions of the type \eqref{eq:Taylor_expansion} provide only a more limited set of physical parameters, for example, the EoS. Since the uncertainties in the nuclear matter properties are easily characterized in terms of uncertainties in the characteristics entering the expansion \eqref{eq:Taylor_expansion}, it is important to establish a one-to-one correspondence between the two descriptions.
% -------------------------------------------------------------------
\begin{table}[t]
\tbl{The coefficients of the expansion \eqref{eq:Taylor_expansion} 
  for the  DDME2 parametrization in MeV units, where we show the values of higher-order parameters not defined in \eqref{eq:Taylor_expansion}, see \cite{Margueron2018a}.}
{
\begin{tabular}{cccccccccc}
\hline\hline
\multicolumn{8}{c}{Density expansion} \\
\cline{4-8}$\rho_{\text{sat}}$ [fm$^{-3}$] & $m^\ast_N/m$& &$E_{\text{sat}}$& &$K_{\text{sat}}$&$Q_{\text{sat}}$&$Z_{\text{sat}}$\\
\hline
      {0.152}&{0.57}& &${-16.14}$& &{251.15} &{479}&4448\\
\hline
\multicolumn{8}{c}{Isospin-asymmetry expansion} \\
\cline{4-8}& & &$ E_{\text{sym}}$&$L_{\text{sym}}$&$K_{\text{sym}}$& $Q_{\text{sym}}$& \\
\cline{4-8}
           & & &{32.31}&{51.27} &  $-87$  & 777 &  \\
\hline\hline
\end{tabular}
}
\label{tab:3}
\end{table}          

%-------------------------------------------------------------------    
The five macroscopic characteristics in Eq.~\eqref{eq:Taylor_expansion} together with the preassigned values of saturation density $\rho_{\rm sat}$ and the nucleon Dirac mass $m^\ast_N/m$ uniquely determine the seven adjustable parameters of the CDFs of the DDME2 type. This allows one to generate new parametrizations  which reproduce desired values of the characteristics~\cite{Li2019PhRvC}, especially those that are associated with high-density and large-isospin behavior. As finite nuclei do not probe these regimes the loss of accuracy of the new CDFs in reproducing the binding energies, charge radii and neutron skins of finite nuclei is marginal.

Having such a tool at the disposal one can now proceed to vary individually the characteristics within their acceptable ranges and extract the EoS of dense matter and, consequently, the properties of compact
stars. This allows one to explore the correlation(s) between specific properties of nuclear matter and/or compact stars~\cite{Caibj2017,Li2019PhRvC,Lijj2019,Cai:2020hkk,Li:2020ass,Li:2021thg}. Detailed investigations of the EoS were carried out where the higher-order coefficients of the expansion, specifically, $Q_{\text{sat}}$ and $L_{\text{sym}}$ were varied since their values are weakly constrained by the conventional fitting protocol of
CDF~\cite{Margueron2018a,Margueron2018b,Margueron2019,Zhangnb2018a,Li2019PhRvC}.  It is interesting that the one-to-one mapping described above allows one to {\it predict} the higher-order terms in the expansion \eqref{eq:Taylor_expansion}~\cite{Li2019PhRvC}, which are highly model dependent~\cite{Dutra2012,Dutra2014}.

\section{Hypernuclear stars: static properties}
\label{sec:HNS_properties}

\subsection{EoS and composition}

Given a CDF one could compute the EoS of the stellar matter by implementing the additional conditions of weak equilibrium and change neutrality that prevail in compact star matter (except for the first instances after birth). The strangeness changing weak equilibrium conditions are given 
%-----------------------------------------------------------------                                     
\begin{eqnarray}
\label{eq:c1}
  &&\mu_{\Lambda}=\mu_{\Sigma^0}=\mu_{\Xi^0}=\mu_{\Delta^0}=\mu_n=\mu_B,\\
\label{eq:c2}
 &&  \mu_{\Sigma^-}=\mu_{\Xi^-}=\mu_{\Delta^-}=\mu_B-\mu_Q,\\
\label{eq:c3}
  &&\mu_{\Sigma^+}=\mu_{\Delta^+}=\mu_B+\mu_Q,\\
\label{eq:c4}
 && \mu_{\Delta^{++}}=\mu_B+2\mu_Q,
\end{eqnarray}
% -----------------------------------------------------------------
where $\mu_B$ and $\mu_Q=\mu_p-\mu_n$ are the baryon and charge chemical potentials, $\mu_i$ with $i \in \{\Lambda, \Sigma^{0,\pm}, \Xi^{0,\pm}, \Delta^{0,\pm,++}\}$ are the chemical potentials of the baryons.  The baryonic charge is given by
% -----------------------------------------------------------------
$$n_p+n_{\Sigma^+}+2n_{\Delta^{++}}
+n_{\Delta^{+}}-(n_{\Sigma^-}+n_{\Xi^-}+n_{\Delta^-})=n_Q,$$
% -----------------------------------------------------------------
where $n_i$ are the baryon number densities.
The electrical charge neutrality is then implemented by equating this quantity with the lepton charge
% -----------------------------------------------------------------
\begin{equation}
n_Q = n_e + n_\mu,
\end{equation}
% -----------------------------------------------------------------
where $n_{e,\mu}$ are the number densities of electrons and muons.
%--------------------------------------------------------
\begin{figure}[hbt]
 \begin{center}
   \includegraphics[width=9.5cm]{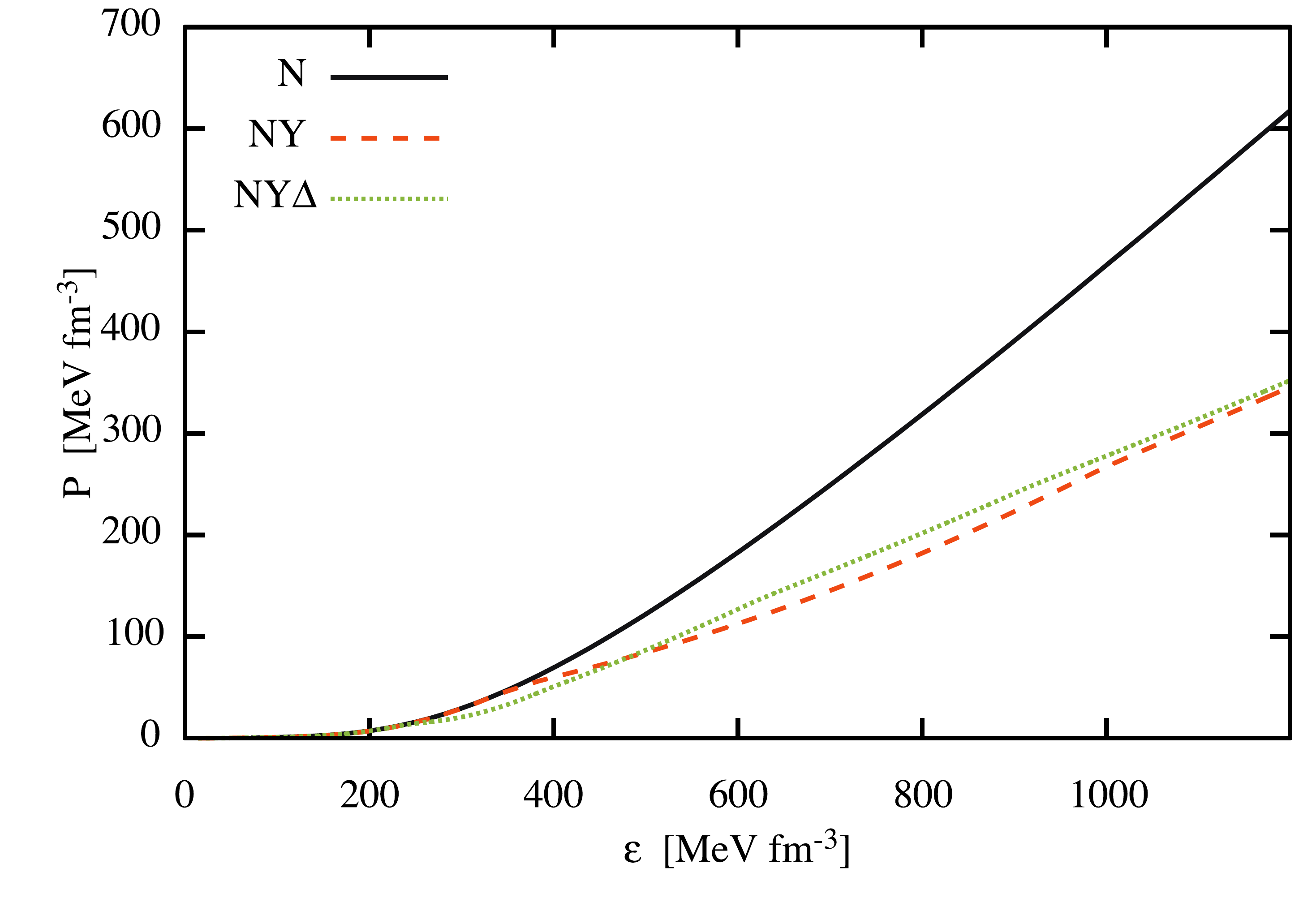} \end{center}
 \vskip -0.5cm
\caption{
  The EoS of nucleonic ($N$), hyperonic ($NY$) and $\Delta$-admixed hyperonic ($NY\Delta$) matter at zero temperature and in $\beta$-equilibrium.  The $\Delta$-potential is fixed by $V_\Delta(\rho_{\rm sat}) = V_N(\rho_{\rm sat})$.}
\label{fig:1.4}
\end{figure}
% --------------------------------------------------------
Figure~\ref{fig:1.4} shows the EoS for nucleonic ($N$), hyperonic ($NY$) and $\Delta$-admixed hyperonic ($NY\Delta$) matter at zero temperature and in weak $\beta$-equilibrium.  The onset of hyperons and $\Delta$ particles are seen by the change in the slope of the pressure above the saturation density and significant softening of the EoS. In the case where $\Delta$'s are included in the composition, the EoS  softens at low and stiffens at high densities compared to the purely hyperonic case.  
\begin{figure}[hbt]
 \begin{center}
   \includegraphics[width=9.5cm]{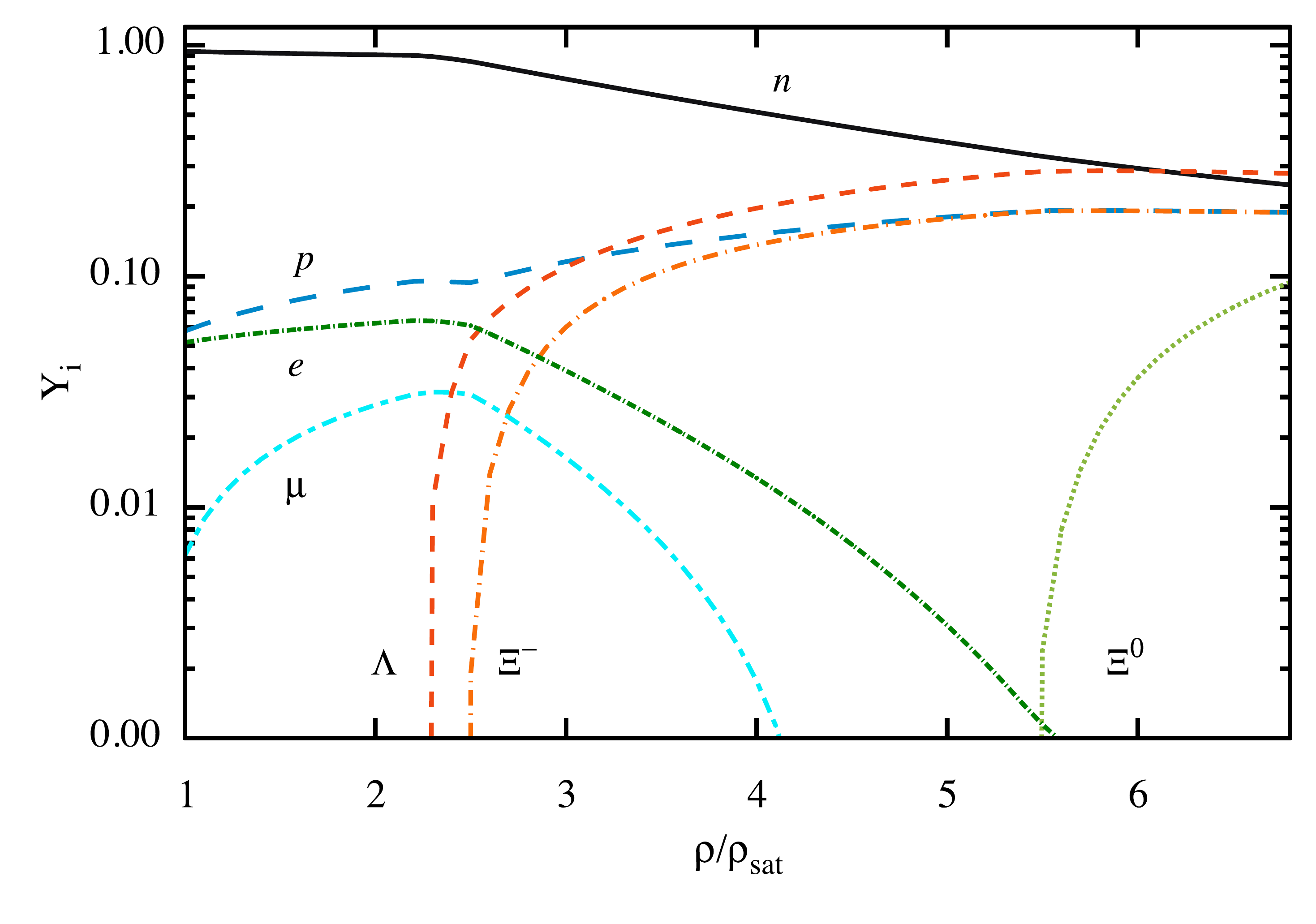}
 \end{center}
 \vskip -0.5cm
 \caption{Particle fractions $Y_i = n_i/n$, where $n_i$ is the number density of species $i$ and $n$ is the
   total density, in zero-temperature hypernuclear matter 
  according to the DDME2 model as a function of the ratio
  $\rho/\rho_{\rm sat}$. }
\label{fig:1.5}
\end{figure}
%--------------------------------------------------------
Figure~\ref{fig:1.5} shows the particle abundances of hyperonic matter for the DDME2 plus hyperons parameterization.  This EoS is smoothly matched at the density $\rho_{\rm sat}/2$ to that of the crust EoS~\cite{Baym1971a,Baym1971b}.  It is seen that the first hyperon to appear is the $\Lambda$, which is followed by the $\Xi^-$ hyperon.  The $\Sigma^-$ hyperons do not appear because they are disfavored by their repulsive potential at nuclear saturation density~\cite{BartPhysRevLett,DOVER1984171,Maslov:2015wba,LopesPhysRevC2014,Gomes:2014aka,Miyatsu:2015kwa}.  Similar ordering of hyperon thresholds was found with other hypernuclear CDFs~\cite{Weissenborn2012a,Weissenborn2012b,Fortin2017,Lijj2018a,Lijj2018b}; note also that this picture is in contrast to the case of free hyperonic gas, in which case $\Sigma^-$ is, in fact, the first hyperon to nucleate~\cite{Ambartsumyan1960SvA}. The modification of the particle abundances in the cases when $\Delta$'s are included is as follows~\cite{Lijj2018b}. For strong enough $\Delta$ potential $V^{(N)}_\Delta$ the $\Delta$-threshold density is considerably lower than that for the $\Lambda$ hyperon.  The larger $V^{(N)}_\Delta$ the lower is the onset densities of $\Delta$'s. Because of its negative charge, the first $\Delta$ resonance to appear is the $\Delta^-$, which competes with negatively charged leptons, effectively eliminates the $\Sigma^-$ and shifts the threshold for the $\Xi^-$ to higher densities. The $\Lambda$ hyperon abundance is weakly affected by the $\Delta$'s.  For $V_\Delta \geq V_N$, the remaining $\Delta^{0,+,++}$ resonances also nucleate.  By electric charge neutrality between the baryons and leptons, the appearance of negatively charged $\Delta^-$ and $\Xi^-$ depletes the electron-muon population.  Finally, for large ($V_\Delta \geq V_N$) potentials $\Delta$'s appear already in $1.4M_{\odot}$ compact stars.
%--------------------------------------------------------
\begin{figure}[t]
 \begin{center}
\includegraphics[width=9.5cm]{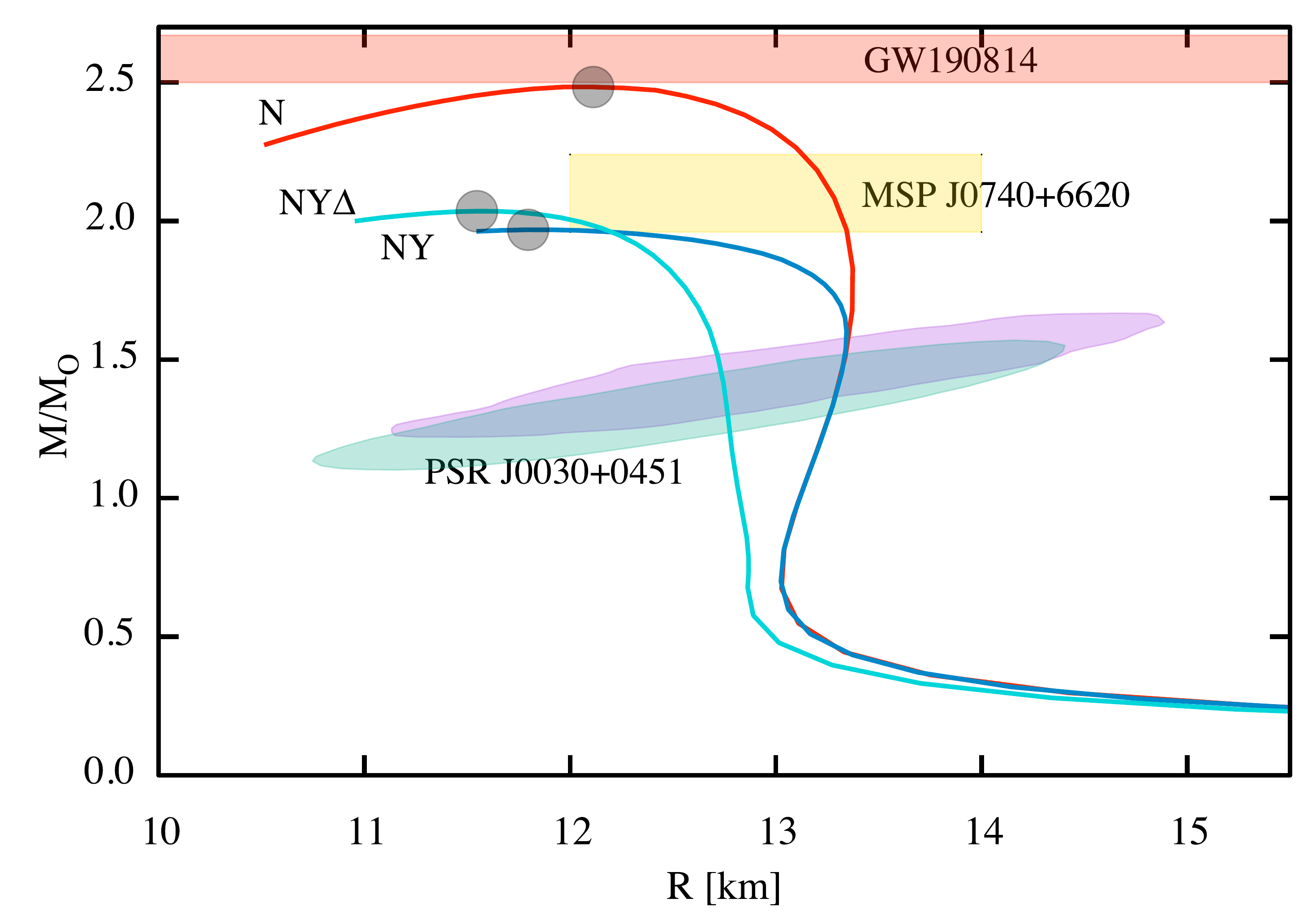} \end{center}
\caption{ MR relations for EoS of state with $N$, $NY$, and $NY\Delta$
  compositions. The shaded areas labeled PSR J0030+0451 corresponds to
  the 1$\sigma$ constraints set by the NICER experiment \cite{Miller2019ApJ,Riley2019ApJ}.
  We also show the 2$\sigma$ range mass for MSPJ0740+6620~\cite{Cromartie2020NatAs}
  and the mass range extracted from the GW190814 event. The maximum mass
  for each stellar sequence is indicated by a solid dot }
\label{fig:1.6}
\end{figure}
% --------------------------------------------------------

\subsection{Global static properties}

The static properties of the compact stars in spherical symmetry (assuming no rotation and significant magnetic fields) are obtained from the integration of the Tolman-Oppenheimer-Volkoff equations~\cite{Tolman1939,Oppenheimer1939}, 
which represent the solution of Einstein's equation for a spherically symmetrical distribution of mass.  It is often useful to compare the theoretical predictions with the observations on plots that contain only observable quantities, i.e. combinations of mass, radius, the moment of inertia, spin frequency, etc. As an example, we show in Fig.~\ref{fig:1.6} the mass-radius (MR) relations for purely nucleonic, hyperonic, and hyperon-$\Delta$ admixed for the EoS presented in Fig.~\ref{fig:1.4}.  The following observational constraints are included: (a) the 1$\sigma$ constraints set by the NICER experiment on the mass {\it and } radius of PSR J0030+0451~\cite{Miller2019ApJ,Riley2019ApJ}; (b) the 2$\sigma$ mass-limit on the largest millisecond pulsar MPS J0740+6620 mass measured~\cite{Cromartie2020NatAs} in combination with the recent NICER teams' report of its 1$\sigma$ radius measurement in the range
$12\le R\le 14$~km~\cite{Miller2021,Riley:2021pdl};  (c) the mass range inferred for the light companion of the binary observed in the GW190814 event to be discussed in detail in Sec.~\ref{sec:Rapid_rotation}.  The softening of the EoS when the heavy baryons are allowed is reflected in the significant reduction of the maximum mass of compact stars in both cases of only hyperons ($NY$) and hyperons and $\Delta$-resonances ($NY\Delta$).  The additional softening of EoS at intermediate densities in the case when $\Delta$-resonances are allowed results in a shift of the radius of the corresponding configuration to smaller values. Since in this case, the stiffness of the EoS is comparable to that of purely hyperonic matter at high densities (and even exceeds it asymptotically) the values of the maximum masses ($M_{\text{max}} \gtrsim 2.0M_\odot$) are comparable in both cases of pure and $\Delta$-admixed hypernuclear matter, see Fig.~\ref{fig:1.6}.

% --------------------------------------------------------
\begin{figure}[t]
  \begin{center}
      \includegraphics[width=9.5cm,height=5.5cm]{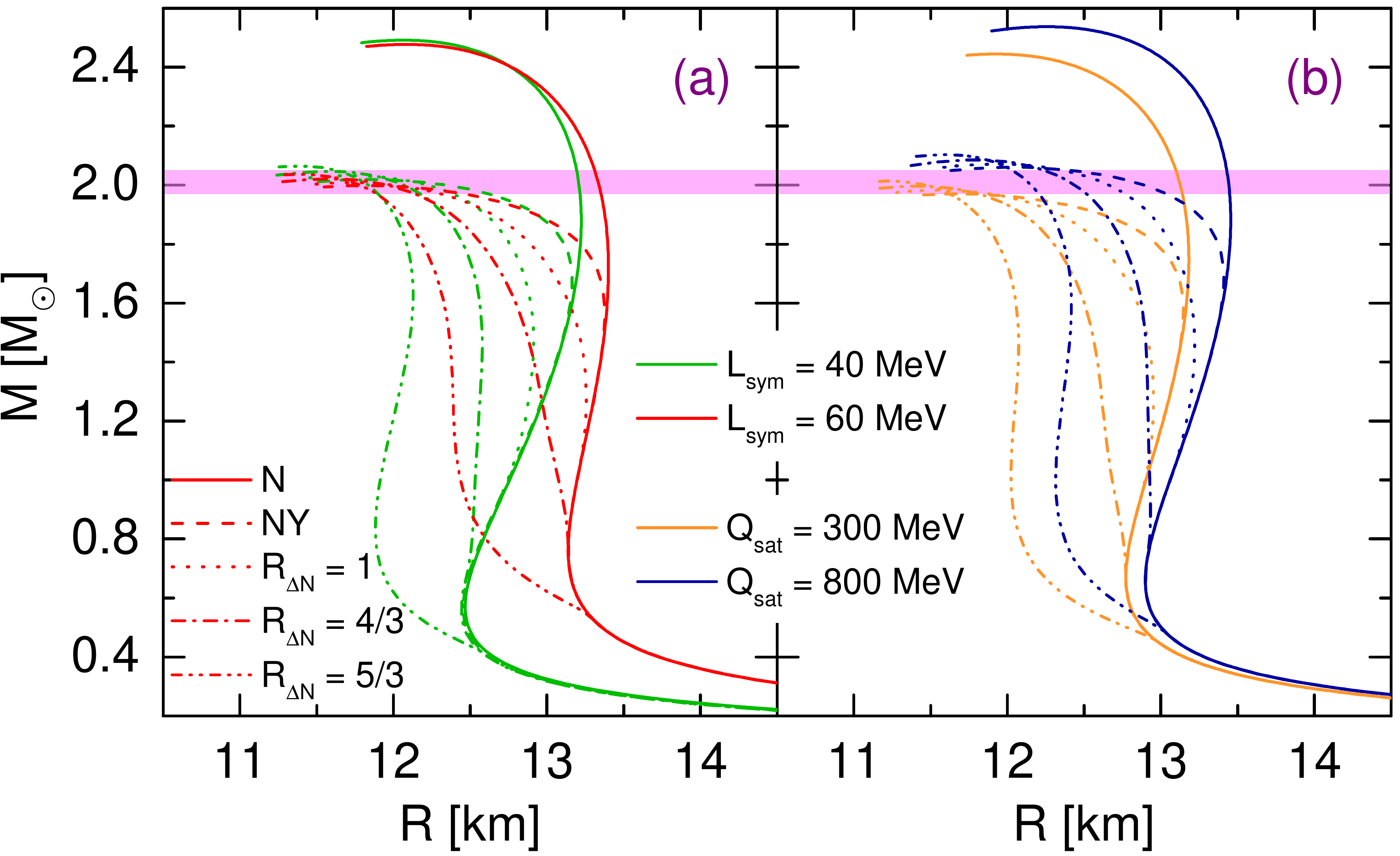}
    \end{center}
    \vskip -0.5cm
\caption{Mass-radius relation for a set of EoS with varying
  $L_{\text{sym}}$ (a) and $Q_{\text{sat}}$ (b) and assuming
  purely nucleonic ($N$), hyperonic ($NY$), and hyperon-$\Delta$
  admixed ($NY\Delta$) compositions of stellar matter \cite{Lijj2019}. Three
  values of the $\Delta$-potential have been used:
  $R_{\Delta N} = V_\Delta/V_N =1$, 4/3 and 5/3, where $V_N$
  is the nucleon potential in isospin-symmetrical matter at
  saturation density. }
\label{fig:1.7}
\end{figure}

%--------------------------------------------------------
To illustrate the effect of variations of the nuclear characteristics on the mass and radius of compact stars Fig.~\ref{fig:1.7} shows the MR plot for the three cases $N$, $NY$ and $NY\Delta$ where the isoscalar $Q_{\text{sat}}$, and the isovector $L_{\text{sym}}$ characteristics are varied, while the remaining parameters in Eq.~\eqref{eq:Taylor_expansion} are fixed at their values predicted by the DDME2 functional (see Table~\ref{tab:3}). Note that the variation of $Q_{\text{sat}}$ is achieved by varying the three density-dependent parameters, i.e., these variations do not impact the meson-hyperon and meson-$\Delta$ couplings at nuclear saturation density. It is seen that smaller values of $L_{\text{sym}}$ and $Q_{\text{sat}}$ imply smaller radius for a given mass. At the same time, a smaller value of $Q_{\text{sat}}$ predicts smaller maximal mass, as the high-density asymptotics of Eq.~\eqref{eq:Taylor_expansion} is reduced.  The effect of varying the $\Delta$-meson couplings are shown by using various values of $\Delta$ potential $R_{\Delta N} = V_\Delta(\rho_{\text{sat}}) /V_N(\rho_{\text{sat}}) = 1; 4/3; 5/3 $. It is seen that the larger is the value of $V_\Delta(\rho_{\text{sat}})$ the smaller is the radius of the predicted configuration, as can be anticipated from the discussion of Fig.~\ref{fig:1.6}. Furthermore, the overall trends are rather similar when varying individually the characteristics for $NY$ and $NY\Delta$ matter, as these are related to the properties of nuclear matter itself and not the heavy baryon admixture.  The differences between the two compositions (i.e. $NY$ vs $NY\Delta$) seen in Fig.~\ref{fig:1.7} (e.g. in the radius of a canonical neutron star) arise from the factors that have been discussed in the context of Fig.~\ref{fig:1.6}.  We close by noting that the parameter space used in Fig.~\ref{fig:1.7} implies maximum masses of configurations larger than $2~M_{\odot}$ and radii within the range $12\lesssim R\lesssim 14$~km independent of the composition.
% --------------------------------------------------------
\begin{figure}[tbh]
  \begin{center}
    \includegraphics[width=8.cm,height=6.9cm]{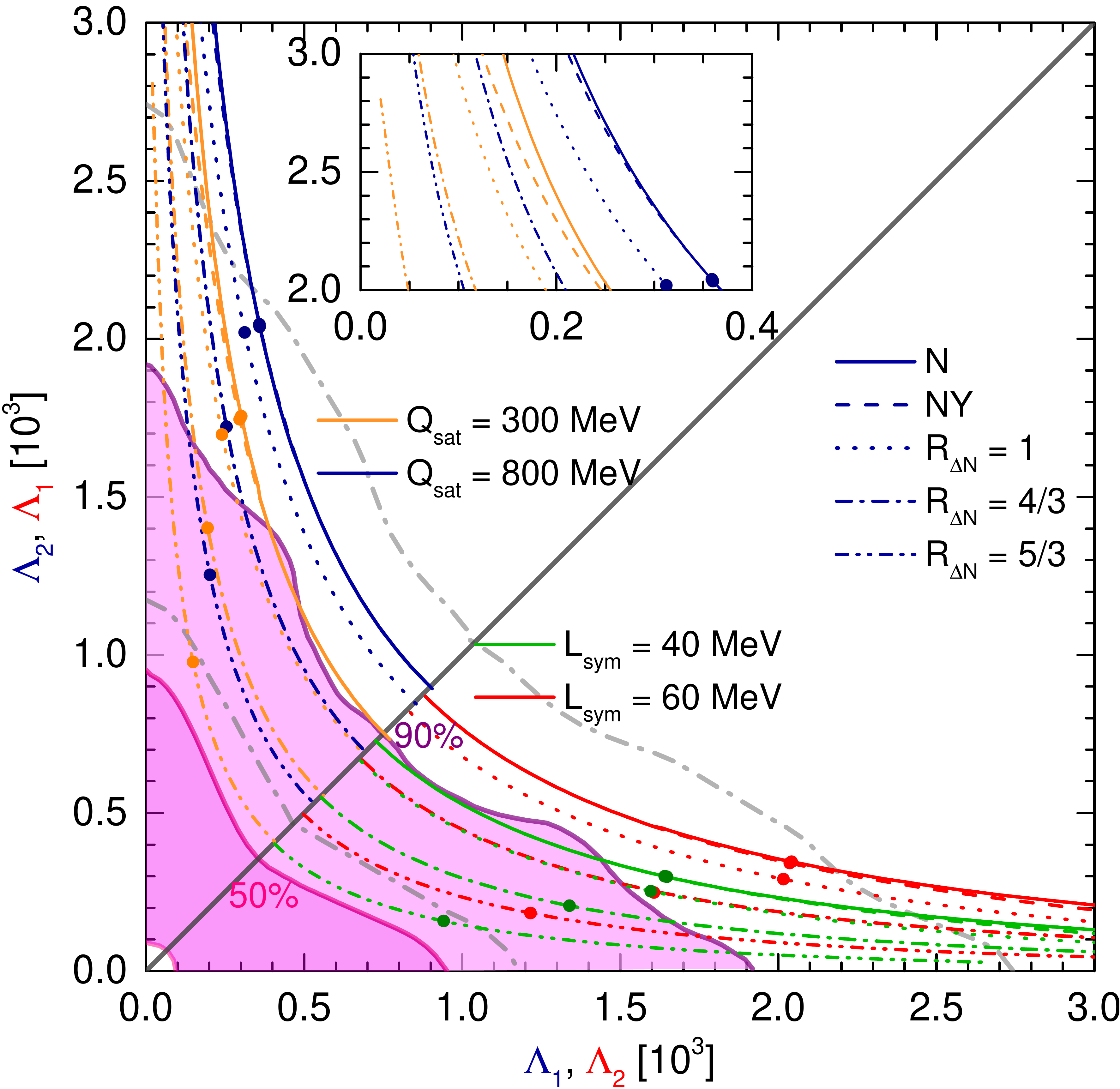}
 \end{center}
 \caption{Dimensionless tidal deformabilities extracted observationally
   from the GW170817 event (shaded areas) are compared with the
   predictions of EoS with various compositions and varying values of 
  $L_{\text{sym}}$, $Q_{\text{sat}}$ and $R_{N\Delta}$ \cite{Lijj2019}.
  The light and heavy shadings correspond to the 50\% and 90\% credibility
  regions \cite{Abbott2019PhysRevX}. The results of the earlier
  analysis by LVC~\cite{Abbott2017a} is shown by  the gray dash-dotted curves. The dots correspond
  to the predictions for a BNS with mass ratio $q = 0.73$. }
\label{fig:1.8}
\end{figure}
% --------------------------------------------------------
\subsection{Tidal deformabilities}

Since the first observation of gravitational waves from a BNS merger - the GW170817 event - the tidal deformability of compact stars has become accessible observationally~\cite{Abbott2017a,Abbott2019PhysRevX} and, thus, can be confronted with the theoretical predictions, as already indicated in Sec.~\ref{sec:Astro_constraints}. Figure~\ref{fig:1.8} shows tidal deformabilities of two stars $\Lambda_1$ and $\Lambda_2$ as defined by Eq.~\eqref{eq:Lambda} for $N$, $NY$, and $NY\Delta$ compositions, three values of the $\Delta$-resonance potential in nuclear matter expressed through the ratio $R_{\Delta N} = V_\Delta/V_N$ and BNS  member masses $M_1$ and $M_2$.  A variation in the characteristics is also allowed by varying $Q_{\text{sat}}$ and $L_{\text{sym}}$.  The diagonal line corresponds to the case of an equal-mass binary assumption for GW170817 event, in which case $M_1 = M_2 = 1.362M_\odot$. The light and heavily shaded areas represent the 90\% and 50\% confidence limits extracted from the analysis of GW170817 event~\cite{Abbott2019PhysRevX}. It is seen that the data favors low values of $Q_{\text{sat}}\simeq 300$~MeV and $L_{\text{sym}}\simeq 40$~MeV for purely nucleonic compact stars, which are otherwise outside the allowed range. The inclusion of hyperons and $\Delta$'s reduces the tidal deformability, as seen clearly from the inset in Fig.~\ref{fig:1.8}. The most significant reduction arises from the inclusion of the $\Delta$'s with large attractive potential, i.e., large $R_{\Delta N}$ value. Thus, the appearance of $\Delta$-resonances with reasonably attractive $\Delta$-potential in the nuclear matter is needed to make the EoS models compatible with the data from GW170817. Purely hyperonic models are compatible with the $90\%$ confidence limits for selected values of $Q_{\text{sat}}$ and $L_{\text{sym}}$.  We note that the softening of EoS at intermediate densities caused by the onset of $\Delta$'s is similar in its effect to a first-order phase transition to a quark matter in which case again more compact configurations arise, see, for example, \cite{Zdunik:2012dj,Bonanno2012,Alford2013,Alford:2017qgh,Alvarez-Castillo:2018pve,Alvarez2019AN,Li:2019fqe,Otto2020EPJST,Fukushima:2020cmk,Kojo2020,Antic2021}. In closing, we note that the information gained from MR and $\Lambda_1-\Lambda_2$ relations are complementary to each other: different EoS models predicting different MR relations could feature $\Lambda_1-\Lambda_2$ relations that are rather close to each other.

\section{Rapidly rotating hyperonic stars}
\label{sec:Rapid_rotation}

A generic feature of Einstein's gravity is the existence of the maximum mass for compact stars assuming that the pressure originates from ``ordinary'' baryonic matter. The exact value of this maximum mass is currently unknown. Rotating compact stars accommodate larger masses than their static (non-rotating) counterparts by about $20\%$ because the centrifugal force provides additional support against the gravitational pull towards the center of the star. There exist several public domain codes for computing stellar configurations of rapidly rotating compact stars, see the RNS code~\cite{Stergioulas2003LRR}  and Rotstar code
~\cite{Gourgoulhon:2021}.  They are based on an iterative method of solution of Einstein's equations~\cite{Nozawa1998,Cook1994} in axial symmetry and use a tabulated EoS as an input.
  The method of solution is iterative: it starts with a ``guess'' configuration (density profile),
  integrates the stellar structure equations, and uses this result as an input for a new iteration.
  This procedure is repeated until convergence to the desired accuracy is reached at each point of the spatial grid.  Rotating compact stars with hyperonic cores have been considered in recent years as well~\cite{Haensel:2016pjp,Lenka2019JPhG}.
% --------------------------------------------------------
\begin{figure}[t]
\begin{center}
\includegraphics[width=9.5cm]{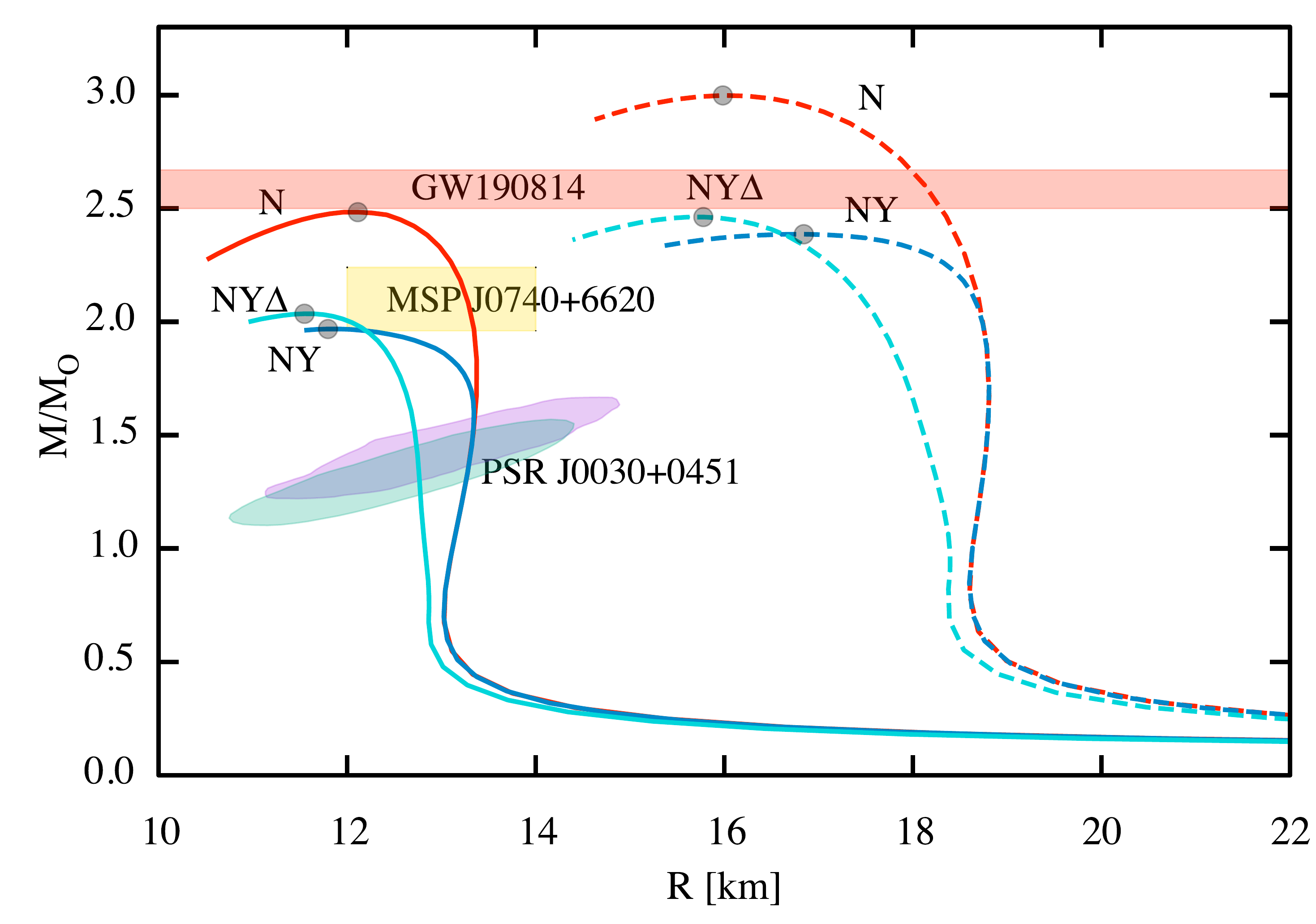} \end{center}
\caption{The mass-radius relations for nonrotating (solid lines) and
  maximally rotating (dashed lines) nucleonic ($N$), hypernuclear
  ($Y$) and $\Delta$-admixed-hypernuclear ($\Delta$) stars.  The
  colored areas show the constraints inferred from the most massive
  pulsar MSP J0740+6620~\cite{Cromartie2020NatAs}, the mass-radius
  limits inferred from the NICER
  experiment~\cite{Miller2019ApJ,Riley2019ApJ} and the mass limits
  from GW190814~\cite{Abbott:2020khf}. The circles indicate the
  maximum masses of the sequences, to the left of which the stars are
  unstable.}
\label{fig:1.13}
\end{figure}
% --------------------------------------------------------
The interest in the exploration of rapidly rotating analogous of the models discussed in the previous sections arose after the measurement by the LVC~\cite{Abbott:2020khf} of gravitational waves from a binary coalescence of a $24.3 M_{\odot}$ black hole with a compact object in the mass the range of $2.50-2.67M_{\odot}$ (the GW190814 event).  The light member of this binary has a mass that lies in the so-called ``mass gap'' $2.5\lesssim M/M_{\odot}\lesssim 5$ where neither a neutron star nor a black hole has been observed and their existence is not obvious from the point of view of stellar evolution scenarios. A natural question that arises in this context is the possible compact star nature of the light companion.  Below we discuss the conjecture that it could be a compact object which is rotating at a frequency that is close to the mass-shedding (Keplerian) limit~\cite{Sedrakian:2020kbi,Li:2020ias} based on an EoS which contains hyperons and $\Delta$-resonances; see also ~\cite{Dexheimer:2020rlp}.
%--------------------------------------------------------

Figure~\ref{fig:1.13} shows the mass-radius relations of compact stars with three compositions $N$, $NY$ and $NY\Delta$ based on the EoS with underlying DDME2 parametrization.  The nucleonic models cover the mass range $2.48\le M/M_{\odot}\le 3$ in the spin frequency range $0\le \Omega \le \Omega_K$, where $\Omega_K$ is the Kepler frequency. Therefore, the nucleonic models account for the mass of a compact star in GW190814 even without rotation.  In the case of $NY$ and $NY\Delta$ compositions, the maximum masses (but not the radii) are quite similar in the static case and this feature extends to the case of rapidly rotating stars.  As already noticed, the softening of the EoS in the case of $NY$ and $NY\Delta$ compositions imply a lower maximum mass compared to the nucleonic case.  The maximum masses for these compositions are close to $2.0M_{\odot}$, therefore the corresponding EoS are inconsistent with the compact star interpretation of the light companion in the GW190814 event.  Their maximally rotating Keplerian analogs on the other hand have maximum masses $\le 2.4M_{\odot}$ suggesting that the maximal rotation is not sufficient to raise masses to the required value $2.5M_{\odot}$. Thus, there exists a significant tension between the hyperonization (with or without an admixture of $\Delta$-resonances) and the interpretation of the light companion of GW190814 as a compact star. The results shown so far were obtained within a specific density functional and a valid question is whether the modifications of the CDF can alter this conclusion.
% --------------------------------------------------------
\begin{figure}[t]
  \centering
    \includegraphics[width=12cm]{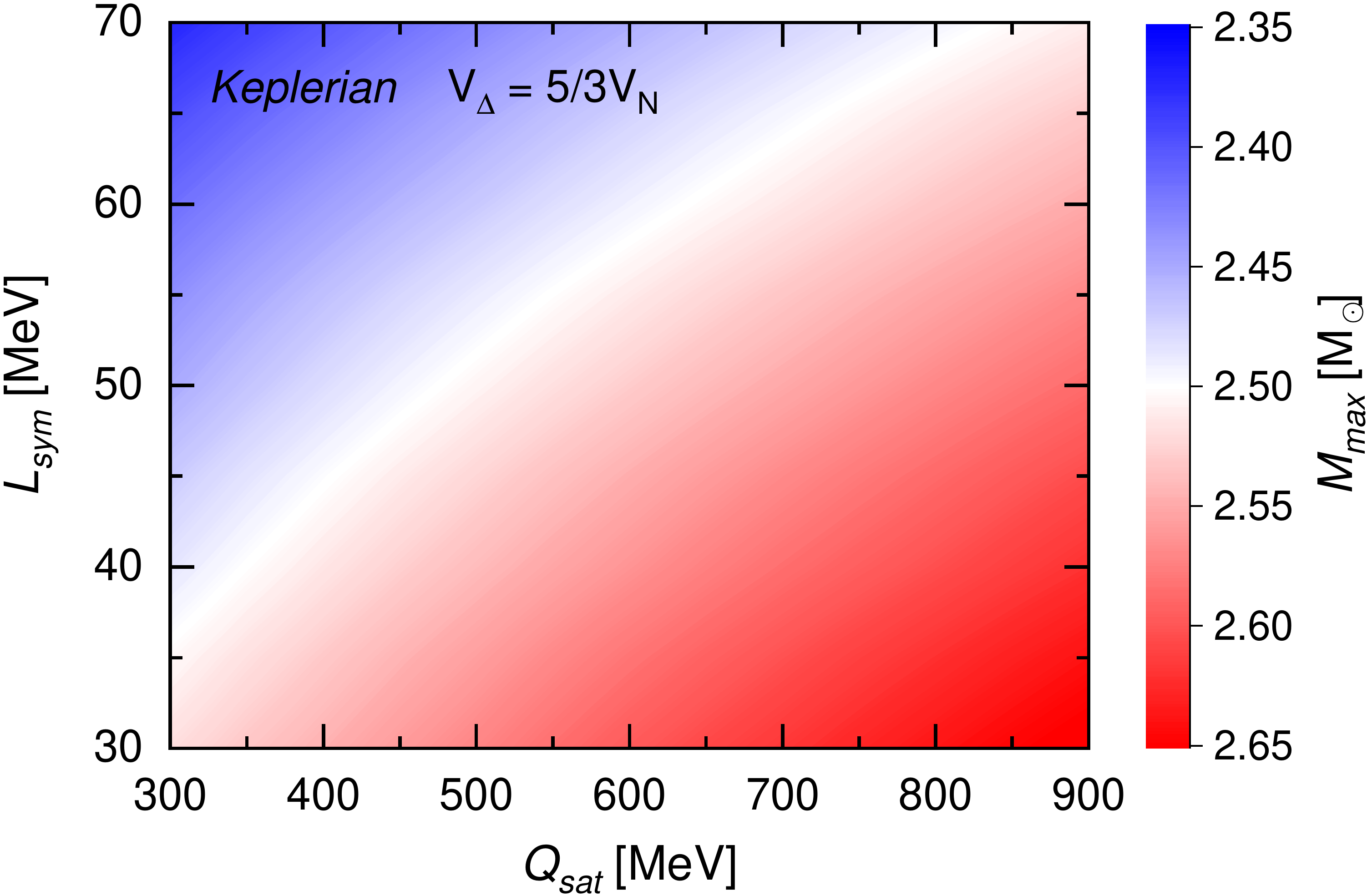}
\caption{The maximum masses of Keplerian sequences
  (color coded column on the right) as a function parameter space
  spanned by $Q_{\rm sat}$ and $L_{\rm sym}$. The $\Delta$-resonance potential is fixed at
  assuming $R_{N\Delta} =5/3$.
  The large-$Q_{\rm sat}$ and small-$L_{\rm sym}$ range corresponds to compact stars
  with masses exceeding $2.5\,M_{\odot}$.}
\label{fig:1.14}
\end{figure}
% --------------------------------------------------------
A study of rapidly rotating stars in the case where the modifications of EoS of $NY\Delta$ matter were framed in the language of the characteristics appearing in Eq.~\eqref{eq:Taylor_expansion}, in particular the values of $Q_{\rm sat}$ and $L_{\rm sym}$ parameters, was carried out in~\cite{Li:2020ias}.  The maximal value $R_{N\Delta}=5/3$ was adopted although, as discussed above, the appearance of $\Delta$-resonances increases the maximum mass of a configuration only slightly.

Figure \ref{fig:1.14} shows the dependence of maximum masses of the Keplerian models on parameters $Q_{\rm sat}$ and $L_{\rm sym}$ for $R_{N\Delta} =5/3$. It is seen that large masses, which are compatible with the compact star in GW190814, arise in the region of large $Q_{\rm sat}$ which imply larger energies at asymptotically large densities. Large masses are also favored for smaller values of parameter $L_{\rm sym}$ (which implies smaller radii of stars and, therefore, more compact objects). One may compare the required values of $Q_{\rm sat}$ with other existing functionals (to avoid, in a first approximation, a full-scale computation with any given functional).  A comparison shows that the range of $Q_{\rm sat}$ values
compatible with a compact star in GW190814 has no overlap with the values predicted by   large samples of non-relativistic and relativistic density functionals~\cite{Dutra2012,Dutra2014}, exceptions
being the DDME2~\cite{Lalazissis2005} and some of the recently proposed functionals~\cite{Taninah:2019cku,Fattoyev2020}.

One may therefore conclude that the secondary object in GW190814 event must have been a low-mass black hole. The compact star interpretation is in strong tension with the idea of hyperonization of dense matter (with and without $\Delta$-resonances). Several extreme assumptions are required to state the contrary, such as maximally rapid rotation, as well as $Q_{\rm sat}$ and $L_{\rm sym}$ values that are outside the range covered by most of the functionals.

\section{{Hyperonization vs clustering at low densities}}
\label{sec:Clusters}

So far we concentrated on the hyperonization in the cold and dense matter as it occurs in the mature compact stars, where the appearance of hyperons was made energetically favorable by the fact that in the compressed matter the neutron Fermi energy can exceed the in-medium rest masses of various hyperons. However, thermodynamic conditions may be favorable for the hyperonization in a different setting, such as in dilute and warm matter that may be (transiently) formed in supernova explosions and BNS mergers~\cite{Nakazato2012,Peres2013}. Because under these conditions nuclear matter is composed of nucleons and clusters (characterized by a mass number and a charge) at certain isospin asymmetry one needs to resort to the ideas of nuclear statistical equilibrium~\cite{Typel2010,Botvina2010NuPhA,Raduta2010,Hempel2012ApJ,Hempel2015PhRvC,Gulminelli2012PhRvC,Gulminelli2015PhRvC,Furusawa2016,Avancini2017,Zhang2019PhRvC,Grams2018PhRvC,Raduta2019,Ropke2020,Pais2020}.  
Additional facets are the pionization (at sufficiently high
temperatures)~\cite{Peres2013,Ishizuka2008,Colucci2014PhLB,Fore2020} and formation of condensates of
deuterons~\cite{Lombardo2001PhRvC,Sedrakian2006PhRvC,Sedrakian:2018ydt} and alpha-particles~\cite{Wu2017JLTP,Zhang2017,Zhang2019PhRvC,Satarov2019PhRvC,Satarov2021PhRvC,Furusawa2020} at low temperatures.

An interesting question to be explored is the interplay between the clustering and heavy-baryon degrees of freedom in dilute, finite-temperature nuclear matter~\cite{Menezes:2017svf,Fortin2018PASA,Sedrakian:2020cjt}. The rich nucleon-$\Delta$-pion dynamics, extensively explored in the heavy-ion context, suggests that one needs to include the quartet of $\Delta$-resonances and the isotriplet of pions $\pi^{\pm,0}$ along with the nucleons and clusters~\cite{Sedrakian:2020cjt}.
In a first approximation, the full nuclear statistical ensemble can be approximated by the light clusters with the mass number $A\le 4$ and a heavy nucleus, for example, $^{56}$Fe. Since the hyperonic interactions are less important at low densities the lightest $\Lambda$-hyperon is expected to contribute dominantly to the strangeness content of matter. 
% --------------------------------------------------------
\begin{figure}[t]
    \includegraphics[width=12cm]{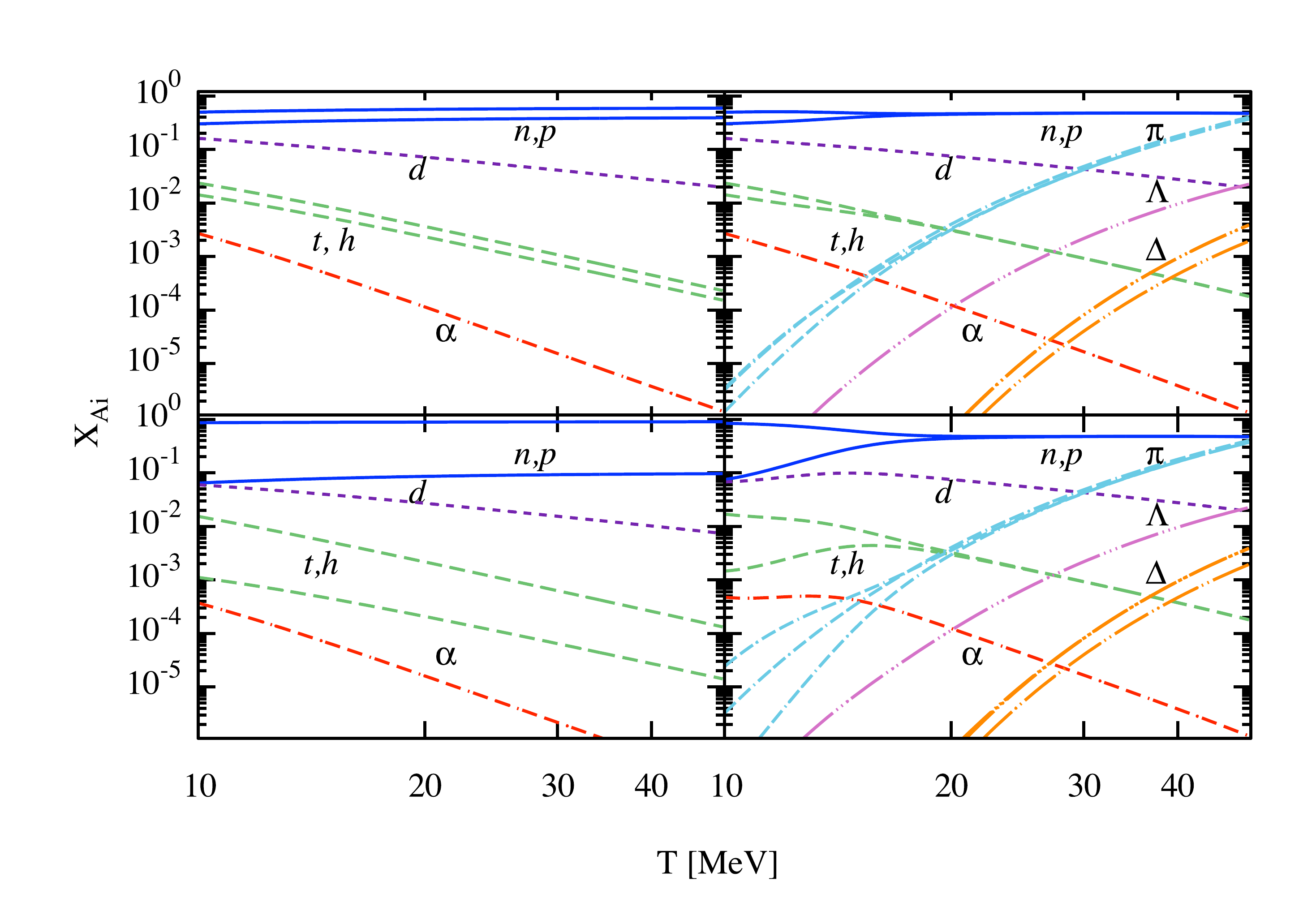}
    \caption{
Mass fractions of the constituents as a function of temperature at 
 fixed density  $10^{-2}\rho_{\rm sat}$ for charge fractions
 $Y_Q = 0.4$ (top panels) and 0.1 (bottom panels). In the two
 left panels include only nucleons and light clusters, whereas the
  two right panels the full composition. The composition includes neutrons $n$, 
  protons $p$, deuterons $d$, triton $t$ helium $h$, $\alpha$-particles, $\Delta$ resonances, $\Lambda$-hyperon and pions $\pi$.  }
\label{fig:1.15}
\end{figure}
% --------------------------------------------------------
As an illustrative example we show in Fig.~\ref{fig:1.15} the composition of dilute nuclear matter in the temperature range $10\le T\le 30$~MeV and density $10^{-2}\rho_{\rm sat}$ for charge fractions $Y_Q = 0.4$ (characteristic for supernova matter) and $Y_Q = 0.1$ (characteristic for BNS mergers)~\cite{Sedrakian:2020cjt}.  We show the {\it mass fraction} $X_j = A_j n_j/n$, where $A_j$ is the mass number of a constituent, $n_j$ is their number density, and $n$ is the total density. The mass fraction of $^{56}$Fe is not visible on the figure's scale.  The main observation is that there is a transition from light baryons and clusters to heavy-baryon featuring matter at about temperature $T_{\rm tr}\sim 30$ MeV. Once heavy baryons and pions are included the isospin asymmetry in the neutron and proton components is reduced. This affects the abundances of the helion and triton which are then closer together. We believe that these are generic,
model-independent features. The transition temperature itself is dependent on the treatment of the interactions and assumed composition; for alternatives see ~\cite{Fortin2018PASA,Menezes:2017svf}. Other generic aspects worth mentioning are: (a) in the presence of a heavy nucleus the abundances of light clusters are strongly suppressed at low temperatures; (b) the phase-space occupation suppresses
completely the light clusters for densities $\ge 0.1\rho_{\rm sat}$ due to the Pauli
blocking~\cite{Ropke2020}; (c) at low enough temperatures deuterons can cross-over from Bose-Einstein to Bardeen-Cooper-Schrieffer pair condensate~\cite{Lombardo2001PhRvC,Sedrakian2006PhRvC,Sedrakian:2018ydt}.

\section{Hypernuclear stars: pairing patterns and cooling}
\label{sec:Cooling} 

The long-term cooling of compact stars is a sensitive probe of the interior composition of such stars~\cite{Shapiro:1983du,SCHAAB1996531,PAGE2006497,Page2009ASS,Sedrakian2007PrPNP,Potekhin2010PhyU}.
It is characterized by neutrino emission from the bulk of the stellar interior during the time-span $t \le 10^5$ yr after the star's birth, which is followed by late-time photon cooling from its surface, assuming that there is no internal heating mechanism operating at any stage of evolution. With the advent of CDFs which were tuned to reproduce the available astrophysical and laboratory data, it became feasible to perform
simulations of cooling of compact stars with hyperonization in a more constrained manner than was possible initially~\cite{Raduta2017,Raduta:2019rsk,Grigorian:2018bvg,Negreiros:2018cho,Fortin:2021umb}.

Hyperonic matter cools via the direct Urca processes~\cite{Prakash:1992zng}. There exist density thresholds
for these processes to operate, which are dictated by the kinematics involved in the reaction, but these densities are very low for hyperons, i.e., they start operating at a density that is slightly above the onset density
for a hyperon participating in a reaction. Baryon pairing is known to suppress the rates of the Urca (and other baryonic) processes, therefore another unknown in the cooling simulations is the magnitude of the gaps in the spectra of various
hyperons~\cite{Raduta2017,Raduta:2019rsk}.

Nevertheless, with the input provided by any given CDF which can be supplemented by the solutions of the BCS equations for the hyperons, the simulations of the cooling of compact stars provided some generic insights~\cite{Raduta2017,Raduta:2019rsk}. There appears to be a mass hierarchy with respect to the cooling behavior. The lightest stars that contain hyperons with $M/M_{\odot}\gtrsim 1.5 $ cool via the Urca process $\Lambda\to p + l +\bar\nu_l$, where $l$ stands for lepton, $\bar\nu_l$ for the associated antineutrino. The appearance of $\Xi^-$ in slightly more massive stars opens a competing channel via $\Xi^-\to \Lambda + l +\bar\nu_l$, the degree of its efficacy depending on the pairing gaps of $\Xi^-$.  For very massive stars with $M/M_{\odot}\sim 2 $ the $S$-wave gaps of protons and $\Lambda$'s will vanish at high densities (because the interactions will become repulsive at large energies) and therefore the Urca process involving $\Lambda,p$ baryons will be the dominant one. Simulations show that the massive is the star the faster it cools. If $\Sigma^-$ appears in the matter, the corresponding Urca process via $\Lambda$ hyperon may be the dominant cooling agent, since $\Sigma^-$ interaction is likely repulsive with no BCS gap in their spectrum. Given the mass hierarchy, the observations of the surface temperatures of neutron stars can be explained by the variation of their masses  within their population (the light objects are hot, the heavier ones -- cold). Note that the massive models may not develop normal cores of hyperons due to the possible pairing in the $P$-wave channel in the high-density matter in which case fast cooling will not take place~\cite{Raduta:2019rsk}. The pairing in the hyperonic sector remains the main unknown for cooling simulations of hypernuclear stars. Given this uncertainty, some studies neglect the hyperonic pairing altogether~\cite{Grigorian:2018bvg,Negreiros:2018cho}.

There are uncertainties in the studies of cooling of neutron stars, that are unrelated to the hyperonic component, which we list here for completeness. These include the composition of the atmosphere~\cite{Potekhin:2020ttj} which substantially affects the surface temperature of the star and the pairing gaps of neutrons and protons in the domains where interactions are attractive~\cite{Sedrakian:2018ydt}. Large magnetic fields are a factor, as they dissipate sufficient energy to heat up the star \cite{Vigano:2021olr}.

\section{{Universal relations}}
\label{sec:Universality}

In recent years universal relations among the global properties of compact stars were established under various conditions, including stationary (non-rotating), rigidly rotating, magnetized, and finite-temperature stars,  for a review, see ~\citet{Yagi2017PhR}.  The universality of relations between global parameters refers to their independence of the EoS that has been input to compute them. These
have proven to be rather useful for the interpretation of observational data because they reduce the uncertainties which are related to the EoS.

In the case of hypernuclear stars (with and without $\Delta$-resonances) it has been established that the universality is maintained among the moment of inertia, quadruple moment and the Love number of the star (the so-called $I$-Love-$Q$ relations) and the dependence of the moment of inertia, quadrupole moment, tidal deformability on the compactness in cold~\cite{Lenka:2017iit} and warm neutron stars~\cite{Raduta2020}. Their extension to finite temperature poses some challenges: it turns out that universalities are maintained 
if one considers stars at the same entropy per baryon $S/A$ and lepton fraction~\cite{Raduta2020}, otherwise the universality is lost~\cite{Maselli:2013mva,Marques:2017zju}.  Clearly, constant entropy approximation will break down under realistic conditions to some degree. While we can be
certain that there are no physical obstacles to maintaining the universality in the finite-temperature domain, a formulation that accounts for entropy gradients needs to be tested.  The universalities so far established extend further to the rapidly rotating stars at finite temperature without loss of accuracy~\cite{Khadkikar:2021yrj}.

\section{{Concluding remarks}}
\label{sec:Conclusions}

The hyperonization of dense matter and the astrophysics of compact stars featuring hyperons and $\Delta$-resonances is a vivid field of research with a strong relation to the observational multimessenger astronomy of compact stars. The recent advances on the observational front, which include the observation of gravitational waves from BNS mergers, notably GW170817, simultaneous measurement of masses and radii of neutron stars and the discovery of very massive pulsars motivate further improvements of the existing models and exploration of their new implications in novel astrophysical scenarios. The ongoing and future hypernuclear programs at CERN, BNL, JLab, GSI-FAIR, and J-PARC, and elsewhere will eventually provide us with insights on hypernuclei as they occur in terrestrial laboratories. These can be used to narrow down the parameters of the CDFs and constrain other theoretical models. The results achieved so far indicate that the combination of theoretical work with the observational advances has the potential of revealing the detailed features of the high-density matter found in compact stars in the observable future.

\section*{{Acknowledgements}}

We are grateful to M. Alford, A. Harutyunyan, M. Oertel, A. Raduta, M. Sinha, V. Thapa for collaboration and insights on the topics related to this chapter. A. S. acknowledges the support by the Deutsche Forschungsgemeinschaft (Grant No. SE 1836/5-1) and the European COST Action CA16214 PHAROS ``The multi-messenger physics and astrophysics of neutron stars''.  F. W. is supported through the U.S. National Science Foundation under Grants PHY-1714068 and PHY-2012152.

%\bibliographystyle{ws-book-har.bst}
%\bibliography{Hyperons_ref.bib,fullref.bib}

\end{document}